\documentclass[aps,pre,reprint,10pt,twocolumn,showpacs]{revtex4-1}
\usepackage{graphicx}
\usepackage{amssymb}
\usepackage{amsmath}
\usepackage{amsfonts}
\usepackage{color}
\usepackage[latin1]{inputenc}
\usepackage{float}
\usepackage{comment}

\newcommand{\expct}[1]{\langle{#1}\rangle}

\begin{document}
\title{Width and extremal height distributions of fluctuating interfaces with window boundary conditions}

\author{I. S. S. Carrasco}
\email{ismael.carrasco@ufv.br}
\author{T. J. Oliveira}
\email{tiago@ufv.br}
\affiliation{Departamento de F\'isica, Universidade Federal de Vi\c cosa, 36570-900, Vi\c cosa, Minas Gerais, Brazil}

\begin{abstract}
We present a detailed study of squared local roughness (SLRDs) and local extremal height distributions (LEHDs), calculated in windows of lateral size $l$, for interfaces in several universality classes, in substrate dimensions $d_s = 1$ and $d_s = 2$. We show that their cumulants follow a Family-Vicsek type scaling, and, at early times, when $\xi \ll l$ ($\xi$ is the correlation length), the rescaled SLRDs are given by log-normal distributions, with their $n$th cumulant scaling as $(\xi/l)^{(n-1)d_s}$. This give rise to an interesting temporal scaling for such cumulants $\left\langle w_n \right\rangle_c \sim t^{\gamma_n}$, with $\gamma_n = 2 n \beta + {(n-1)d_s}/{z} = \left[ 2 n + {(n-1)d_s}/{\alpha} \right] \beta$. This scaling is analytically proved for the Edwards-Wilkinson (EW) and Random Deposition interfaces, and numerically confirmed for other classes. In general, it is featured by small corrections and, thus, it yields exponents $\gamma_n$'s (and, consequently, $\alpha$, $\beta$ and $z$) in nice agreement with their respective universality class. Thus, it is an useful framework for numerical and experimental investigations, where it is, usually, hard to estimate the dynamic $z$ and mainly the (global) roughness $\alpha$ exponents. The stationary (for $\xi \gg l$) SLRDs and LEHDs of Kardar-Parisi-Zhang (KPZ) class are also investigated and, for some models, strong finite-size corrections are found. However, we demonstrate that good evidences of their universality can be obtained through successive extrapolations of their cumulant ratios for long times and large $l$'s. We also show that SLRDs and LEHDs are the same for flat and curved KPZ interfaces.
\end{abstract}


\maketitle

\section{Introduction}
\label{secintro}

The Kardar-Parisi-Zhang (KPZ) equation~\cite{kpz}
\begin{equation}
\frac{\partial h(\vec{x},t)}{\partial t} = \nu \nabla^2 h + \frac{\lambda}{2} (\nabla h)^2 + \eta
\label{eqKPZ}
\end{equation}
is a paradigmatic model in out-of-equilibrium statistical physics. Originally proposed to describe growing interfaces, $h(\vec{x},t)$ can be viewed as the height at substrate position $\vec{x}$ and time $t$, $\nu$ as a surface tension, $\lambda$ as a ``velocity excess'' and $\eta(x,t)$ as a white noise with $\expct{\eta(x,t)} = 0$ and $\expct{\eta(x,t)\eta(x',t')} = 2D\delta^{d_s}(x-x')\delta(t-t')$~\cite{barabasi}. However, fluctuations in diverse other systems as, for example, one-dimensional driven lattice gases~\cite{Kriecherbauer.Krug-JPA2010,*Corwin-RMTA2012} and free fermions in a harmonic well \cite{Dean15}, directed polymers in a random media \cite{healy95} and confined ledges of crystalline facets \cite{Ferrari04,*einstein14} are also described by Eq. \ref{eqKPZ}.

Since 2010, there has been a renewed interest on KPZ systems, mainly due to theoretical calculations of height distributions (HDs) in one-dimension $(d=1+1)$~\cite{SasaSpo1,*Amir,*Calabrese,*Imamura}, and reliable experimental realizations of this class in $d=1+1$~\cite{TakeuchiPRL,*TakeuchiSP,yunker}. In short, it is now know that the $1+1$ growth regime KPZ (HDs) are given by Tracy-Widom (TW) distributions~\cite{tw} from different ensembles depending on surface geometry (flat or curved), whereas the stationary HD is the Baik-Rains \cite{Baik} distribution. Moreover, the temporal and spatial correlators are also different for flat and curved KPZ interfaces~\cite{PraSpo3,*Sasa2005,*Borodin2008}. Extensive numerical simulations have confirmed these results in $d=1+1$ \cite{Alves11,Oliveira12,Alves13,healyCross}, and showed the existence of similar scenarios in $2+1$ ~\cite{healy12,Oliveira13,healy13,Ismael14} - where experimental evidences of universal HDs have been given in~\cite{Almeida14,healy14,Almeida15} - and higher dimensions~\cite{Alves14}.

Beyond the HDs, other fluctuations at surface can present universality. In 1994, R\'acz and coworkers \cite{racz1,racz2,racz3} demonstrated that global squared width (or roughness) distributions (WDs) - calculated at the steady-state regime with periodic boundary conditions (PBC) - are universal. Since then, PBC-WDs have been widely applied in numerical studies of growth models \cite{fabioRDs,vladimir08,Kelling,racz6}, being known to present smaller finite-size corrections than HDs and roughness scaling \cite{tiago07Rug}. For linear interfaces, the exact probability density functions (pdf's) of WDs were calculated for PBC \cite{racz1,*racz2,racz4,racz5} and window boundary condition (WBC) \cite{racz5}, in $d=1+1$. The latter - with the squared local roughness (SLR) calculated in windows of lateral size $l$ that span the surface (of size $L > l$) - being of special importance for experimental analysis, where usually it is pretty hard to attain the steady state. Indeed, the comparison of $2+1$ growth regime SLRDs (with WBC) from vapor deposited films \cite{Almeida14,healy14,Almeida15} and the ones for KPZ models \cite{FabioGR,healy14} have provided experimental evidences of their universality. More recently, a similar study in $d=1+1$ \cite{HHTake2015} have led to the same conclusion for the SLRDs of the celebrated KPZ turbulent liquid-crystal (TLC) experiment \cite{TakeuchiPRL,*TakeuchiSP}, where evidences were provided that the $1+1$ KPZ SLRD agree with the Edwards-Wilkinson (EW) class (defined by Eq. \ref{eqKPZ} with $\lambda=0$) one, calculated exactly by Antal \textit{et al.} \cite{racz5}. 

Other interesting measures at surface are the extremal heights - maximal and minimal heights relative to the mean -, which are associated with drastic events such as a short-circuit in a battery or the breakdown of a device due to corrosion \cite{raychaudhuri01}. The fluctuations of the steady state (global) extremal heights (with PBC) have also been studied and universal distributions were found for several universality classes, including the KPZ one \cite{raychaudhuri01,racz7,majumdar04,majumdar06,lee05,tiago08Ext,schehr10}. Local extremal HDs (LEHDs) for KPZ class (with WBC) in $d=2+1$ was initially studied in \cite{Almeida14}, being very important to support the KPZ universality of CdTe/Si(001) films. Short after, a similar study was done for oligomer films \cite{healy14}, where a more detailed numerical study - establishing the universality of these distributions in KPZ class - was presented. Finally, the same analysis has been applied for the $1+1$ TLC interfaces and, beyond its universality, evidences was found \cite{HHTake2015} that the global extremal HD for EW surfaces with free BC \cite{majumdar04} plays also the role for KPZ with WBC (in $1+1$). 

SLRDs and LEHDs (as well as HDs) have also been recently calculated for electrodeposited oxide films \cite{Iuri15}, providing strong evidences of diffusion dominated growth - Mullings-Herring (MH) class \cite{Mullins,*Herring} - in these systems. Despite all these applications \cite{Almeida14,healy14,Almeida15,HHTake2015,Iuri15}, some aspects of the local distributions remains unexplored as, for example, their short time regime. Moreover, the role of finite-size corrections in such distributions needs more analysis, as pointed out very recently in \cite{FabioGR2}. Here, we present a thorough numerical/theoretical analysis of these aspects considering models in KPZ and other universality classes. We show that the best way to access the universality of these local distributions is through successive extrapolations of their cumulant ratios in time and size, since they present strong $l$-dependence in some systems. Although HDs and correlators are different for the (full) flat and curved interfaces \cite{SasaSpo1,*Amir,*Calabrese,*Imamura,PraSpo3,*Sasa2005,*Borodin2008,healy12,Oliveira13,healy13,Ismael14}, we find equal WBC distributions in both geometries. Interestingly, at short times, the cumulants of SLRDs evolves in time following scaling relations that allow us to determine the (global) scaling exponents of roughening systems from a local, growth regime measure.

The rest of this paper is organized as follows. In Sec.~\ref{secModels}, we define the studied models and the growth methods, as well as the quantities of interest in this work. The scaling of the cumulants of the distributions is presented in Sec.~\ref{secScaling}. In sections~\ref{secSLRDs} 
and~\ref{secLEHDs} the universality of SLRDs and LEHDs, respectively, is analyzed. Our final discussions and conclusions are summarized in Sec. \ref{secConcl}.

\section{Models and methods}
\label{secModels}

\subsection{Models}

Most of the results presented in the following sections are for the (KPZ) restricted solid-on-solid (RSOS)~\cite{kk}, the Etching~\cite{Mello} and the single step (SS)~\cite{barabasi} models, grown on flat substrates with fixed and enlarging sizes. However, in Sec. \ref{secScaling}, we also study the random deposition (RD), the Family \cite{Family} (EW class), the conserved RSOS (CRSOS - Villain-Lai-Das Sarma (VLDS) class \cite{Villain,*LDS}) and the large curvature (LCM - MH class) models. In all cases, the interfaces were grow with periodic boundary conditions. In the Monte Carlo simulation of these models, at each time step, a position $i$ of a substrate with $N$ sites is randomly sorted and, then, the height $h_i$ of this site and/or its neighbors change according to the rules:

\textit{RSOS}: $h_i \rightarrow h_i +1$, if the condition $|h_{j}-h_{i}| \leq 1$ is satisfied for all nearest neighbors (NN) $j$ of site $i$. Otherwise, $h_i$ remains unchanged;

\textit{Etching}: first, if $h_j < h_i$, we make $h_j = h_i$ for all NN $j$. Then, $h_i \rightarrow h_i + 1$;

\textit{SS}: $h_i \rightarrow h_i + 2$, if the condition $h_{j}-h_{i} = 1$ is satisfied for all NN $j$. Otherwise, $h_i$ remains unchanged;

\textit{RD}: $h_i \rightarrow h_i + 1$;

\textit{Family}: $h_k \rightarrow h_k + 1$, where $k$ is the position with the minimal height among $i$ and the NN $j$.

\textit{CRSOS}: if the RSOS rule is not fulfilled at site $i$, the particle diffuse on surface until find a site satisfying it;

\textit{LCM}: $h_k \rightarrow h_k + 1$, where $k$ is the position with the largest surface curvature among $i$ and the NN $j$.

After each deposition \textit{attempt} the time is increased by $\Delta t=1/N$. The initial conditions are (chessboard) $h_{i}$ alternating between 0 and 1 for the SS model, and (flat) $h_i=0$ for the other models.

The method used for the substrate enlargement consists in duplicating columns - i. e., a column $i$ (or a row $j$, in $d=2+1$) is randomly sorted and, then, an identical column (or row) is created at position $i+1$ (or $j+1$)~\footnote{In the SS model, a pair of neighboring columns has to be duplicated in order to conserve the symmetry of up and down steps at surface} - with a rate $\omega$ in each substrate direction \cite{Ismael14}. These duplications  (occurring with probability $P_a = \omega d_s/(N+\omega d_s)$) are randomly mixed with deposition attempts (which have probability $P_d = N/(N+\omega d_s)$), with $d_s=d-1$. After each event (deposition attempt or column duplication) the time is increased by $\Delta t = 1/(N + \omega d_s) $. We start the growth on substrates of lateral size $L_0=\omega$ and, thus, at the time $t$, its (average) size will be $\left\langle L \right\rangle = L_0 + \omega t$. Since the value of the substrate enlarging rate $\omega > 0$ has negligible effects on universal properties of the interfaces \cite{Ismael14}, we consider here only one value of $\omega$, shown in Tab. \ref{tab0}.

\begin{table}[!b]
\begin{center}
\begin{tabular}{cccccccccc}
\hline\hline
 $d$        &  & $\omega$ &  & $t_{max}$ &  &  substrate size   &  &  samples  &   \\
\hline
$(1+1)$     &  &    0     &  &  $2\times 10^4$   &  &  $L = 2^{17}$                   &  &  $3000$   &   \\
$(1+1)$     &  &   20     &  &  $2\times 10^4$   &  &  $\left\langle L \right\rangle_{max} \approx 4 \times 10^5$    &  &  $3000$   &   \\
\hline
$(2+1)$     &  &    0     &  &  $10^3$           &  &  $L = 2^{11}$                   &  &  $1500$   &   \\
$(2+1)$     &  &    4     &  &  $10^3$           &  &  $\left\langle L \right\rangle_{max} \approx 4000$             &  &  $1500$   &   \\
\hline\hline
\end{tabular}
\caption{Details of the simulations of flat models. $t_{max}$ is the maximal deposition time, which leads to the maximal average size $\left\langle L \right\rangle_{max} = \omega (t+1)$ for $\omega > 0$.}
\label{tab0}
\end{center}
\end{table}

We also study the version A of the (KPZ) Eden model~\cite{eden} on the square lattice, where, starting from a single seed at the origin, a radial cluster is grown by randomly adding particles at one of the $N_p$ empty sites at its periphery. In order to eliminate the lattice imposed anisotropy, the growth on a given site $i$ happens with probability $p_i=(n_{i}/4)^{\kappa}$, where $n_i=1,2,3$ or $4$ is the number of occupied nearest neighbors of site $i$ and $\kappa$ is a parameter set to $\kappa=1.705$~\cite{paiva07}. When a particle is added, the time is increased by $1/N_p$. We run averages over 2500 clusters with up to $5.5 \times 10^9$ particles.

\subsection{Quantities of interest}

The squared local roughness at position $i$ of a given surface at the time $t$ is defined as the variance of the heights of the $l^{d_s}$ sites inside a box of lateral size $l$, centered at site $i$
\begin{equation}
w_{2,i}(l,t) \equiv \bar{h^2}_{i}(l,t) - \bar{h}_{i}(l,t)^2.
\end{equation}
Measuring $w_{2,i}(l,t)$ for all positions $i$ of the substrate of size $L \geq l$, as well as for different surfaces (samples) at the same time $t$, we obtain a set of values and from them we built the squared local roughness distributions (SLRDs) $P(w_2)$, so that $P(w_2)dw_2$ gives the probability of finding the squared roughness in the range $[w_2, w_2 + dw_2]$. In a similar way, we define the maximal ($M$) and the minimal ($m$) relative height within a given box $i$, respectively, as
\begin{equation}
M_i \equiv h_{max,i}- \bar{h}_{i} \quad \text{and} \quad m_i \equiv |h_{min,i} - \bar{h}_{i}|,
\end{equation}
where $h_{max,i}$ ($h_{min,i}$) is the maximal (minimal) height inside the box $i$. From the values of $M_i$ and $m_i$ for different $i$'s and samples, we built the maximal [$P(M)$ - MAHDs] and the minimal [$P(m)$ - MIHDs] height distributions, respectively, so that $P(X)dX$ is the probability of finding $X$ in the interval $[X,X+dX]$, with $X=M$ or $m$.

At first glance, the probability density functions (pdf's) $P(X)$ (with $X=w_2, M$ or $m$) may depend on the box size $l$, the time $t$ and parameters related to the specific growing system. So, we must compare rescaled distributions and the best way to do this is making \cite{tiago07Rug,racz7}
\begin{equation}
P(X)=\frac{1}{\sigma_{X}} \Phi_X \left( \frac{X- \left\langle X \right\rangle }{\sigma_{X}}\right),
\label{eqResc}
\end{equation}
where $\sigma_{X}\equiv\left\langle X^2 \right\rangle_c^{1/2}$ is the standard deviation of $X$. The function $\Phi_X$ is expected to assume universal forms for a given class and dimension. In order to characterize this pdf's, we will analyze the first adimensional cumulant ratios of the \textit{non-rescaled} $P(X)$: $R \equiv {\left\langle X \right\rangle}/{\left\langle X^2 \right\rangle_c^{1/2}}$, the skewness $S \equiv {\left\langle X^3 \right\rangle_c}/{\left\langle X^2 \right\rangle_c^{3/2}}$, and the kurtosis
$K \equiv {\left\langle X^4 \right\rangle_c}/{\left\langle X^2 \right\rangle_c^{2}}$, where $\left\langle X^n \right\rangle_c$ is the $n$th cumulant of the fluctuating variable $X$.

\section{Cumulant scaling}
\label{secScaling}

In this section, the time evolution of the local distributions is investigated. All results presented here for \textit{flat} models were obtained for fixed size substrates. We have checked notwithstanding that similar results are found for enlarging systems.

\begin{figure}[!b]
\includegraphics[width=8.1cm]{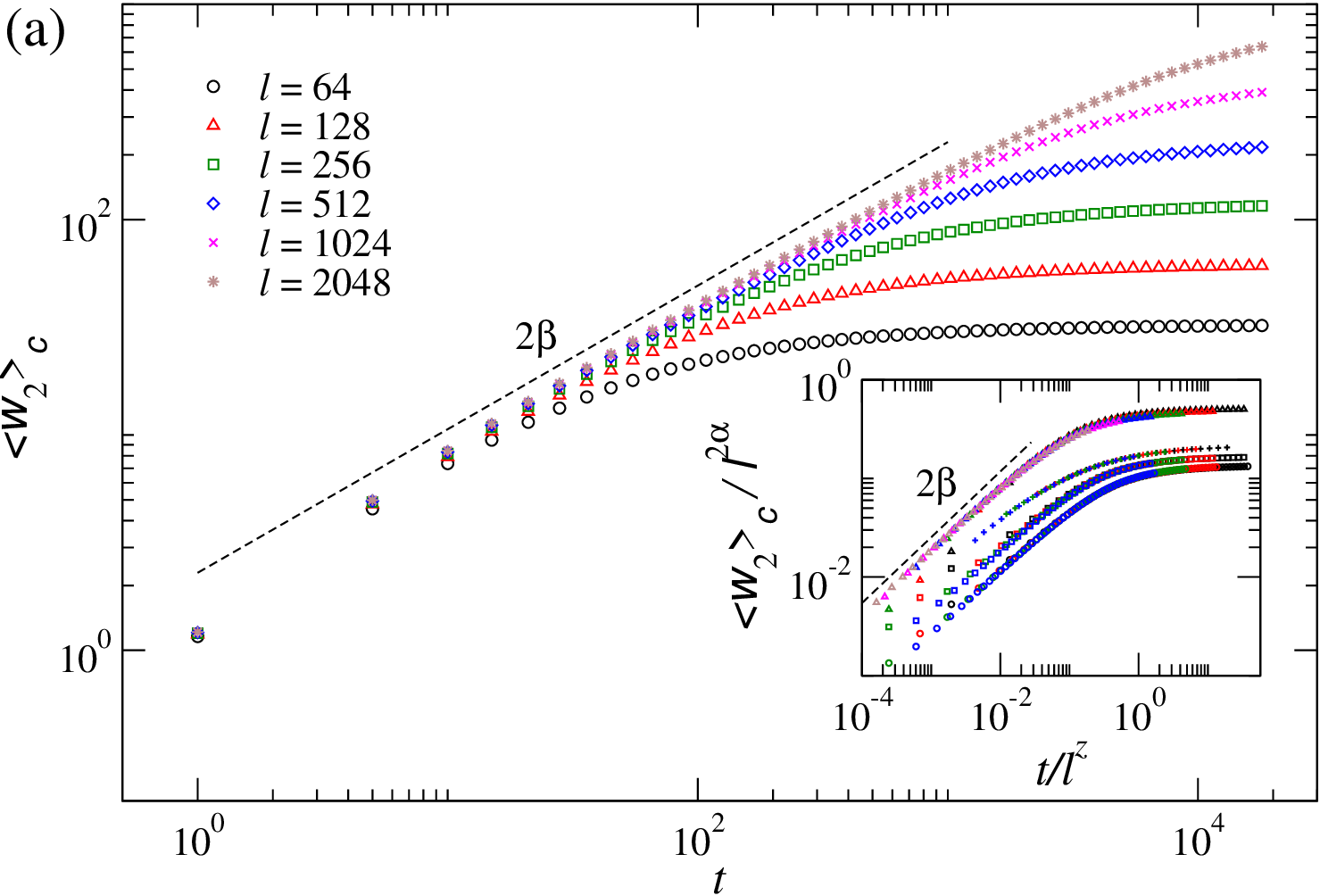}
\includegraphics[width=8.1cm]{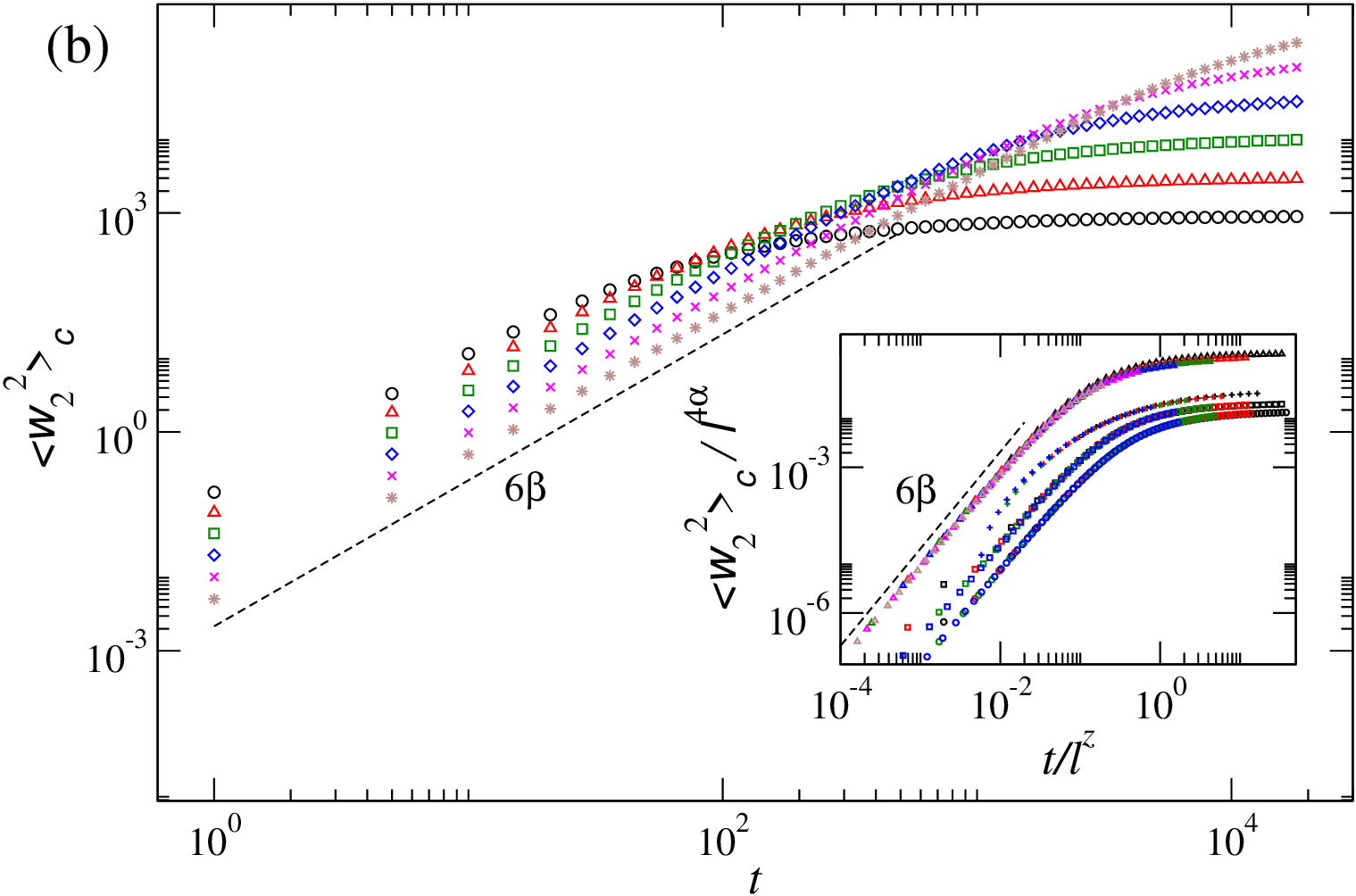}
\includegraphics[width=4.cm]{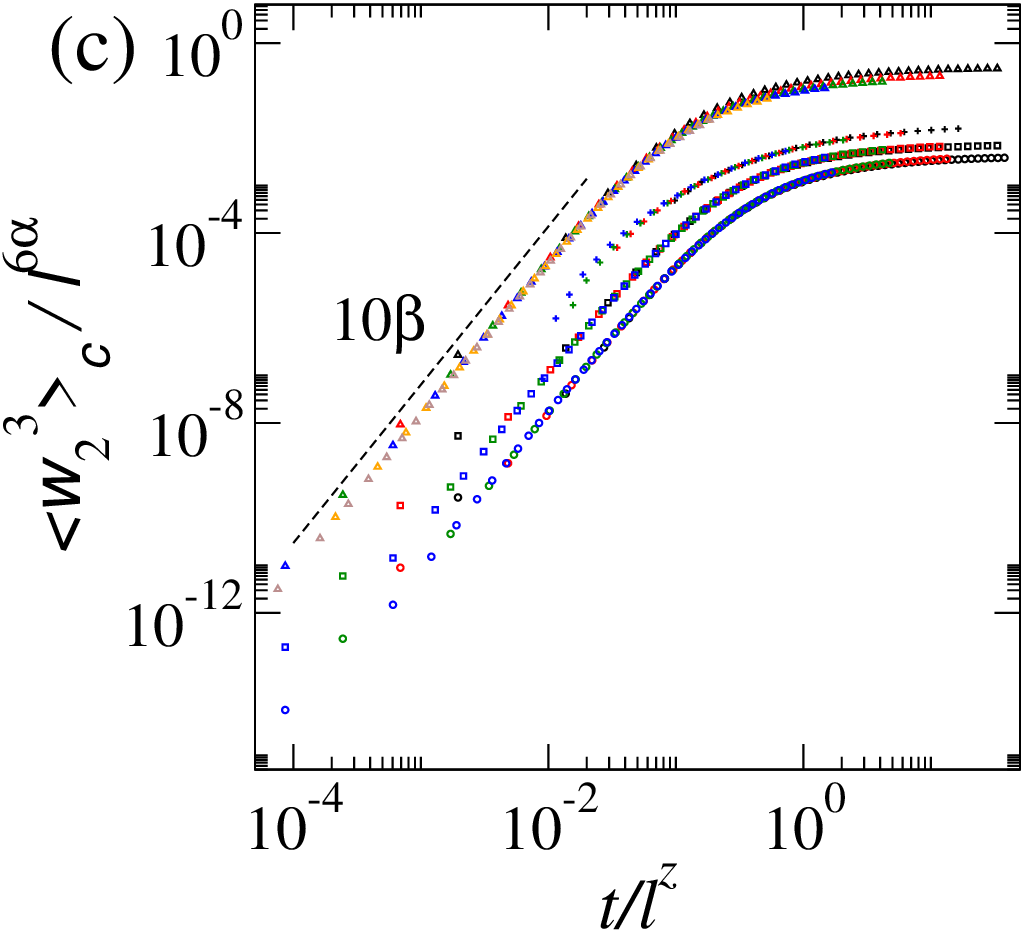}
\includegraphics[width=4.cm]{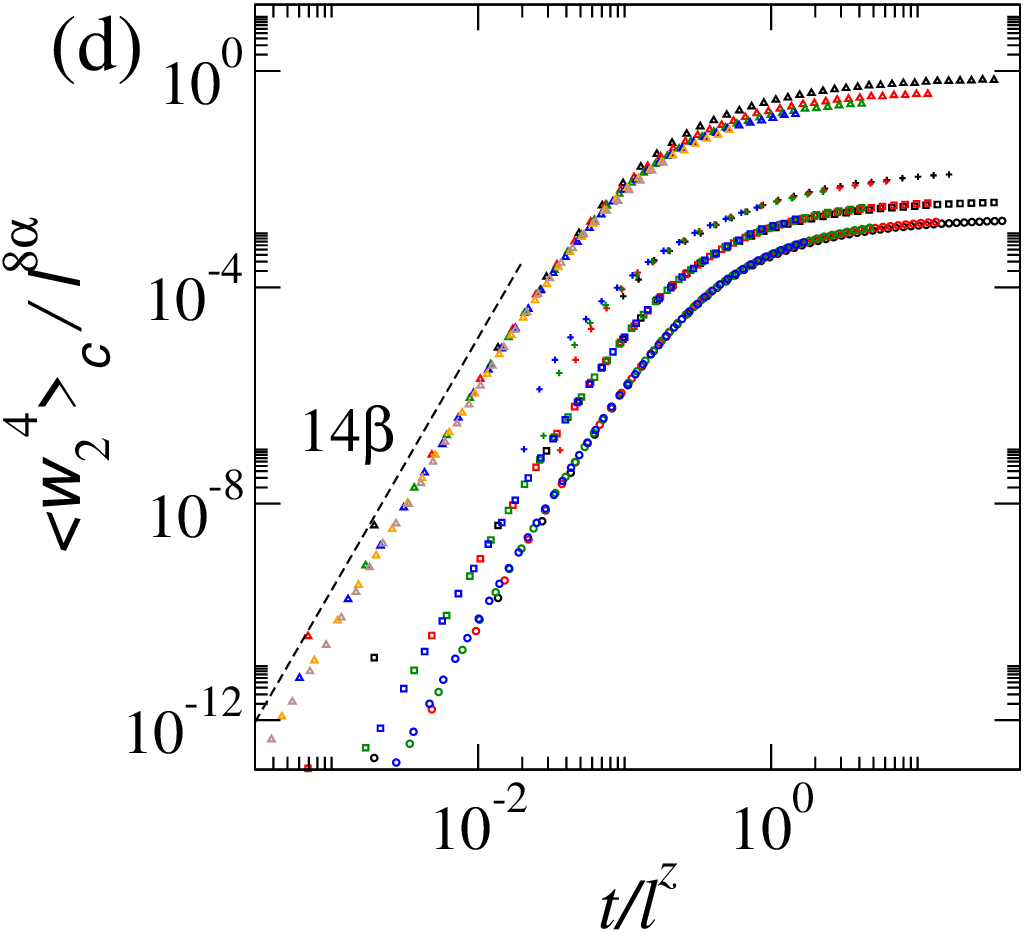}
\caption{(Color online) (a) First and (b) second cumulants of SLRDs as a function of time $t$, calculated in different box sizes $l$ for $d=1+1$ surfaces of the Etching model. The insets of (a) and (b) and the plots in (c) and (d) show the collapse of the rescaled cumulants for all KPZ models in $d=1+1$. From bottom to top, collapsed curves for RSOS, SS, Eden and Etching models are shown. The dashed lines have the indicated slopes, where $\beta=1/3$.}
\label{fig1}
\end{figure}

\subsection{SLRDs}

The first cumulant (the mean) of the SLRDs, $\left\langle w_2 \right\rangle_c$, is the squared local roughness - a standard quantity in the analysis of fluctuating interfaces \cite{KrugAdv}. In KPZ systems, the scaling of $\left\langle w_2 \right\rangle_c$ in time $t$ and box size $l$ is given by the Family-Vicsek (FV) \cite{FV} scaling $\left\langle w_2 \right\rangle_c \sim l^{2 \alpha} f_1\left( t/l^{z}\right)$, where $f_1(x)$ is a scaling function behaving as $f_1(x) \approx const.$ for $x \gg 1$ and $f_1(x) \sim x^{2 \beta}$ for $x \ll 1$, with $\beta=\alpha/z$ being the growth exponent. Therefore, for small times $(t \ll l^{z})$, so that the lateral correlation length ($\xi \sim t^{1/z}$) is smaller than $l$, $\left\langle w_2 \right\rangle_c \sim t^{2\beta}$. On the other hand, for $t \gg l^{z}$ (i. e, for $\xi \gg l$), $\left\langle w_2 \right\rangle_c$ becomes constant in time and scales as $\left\langle w_2 \right\rangle_c \sim l^{2\alpha}$. This well-known scaling is shown in Fig. \ref{fig1}a for the ($1+1$) Etching model.

\begin{figure}[!t]
\includegraphics[height=4.cm]{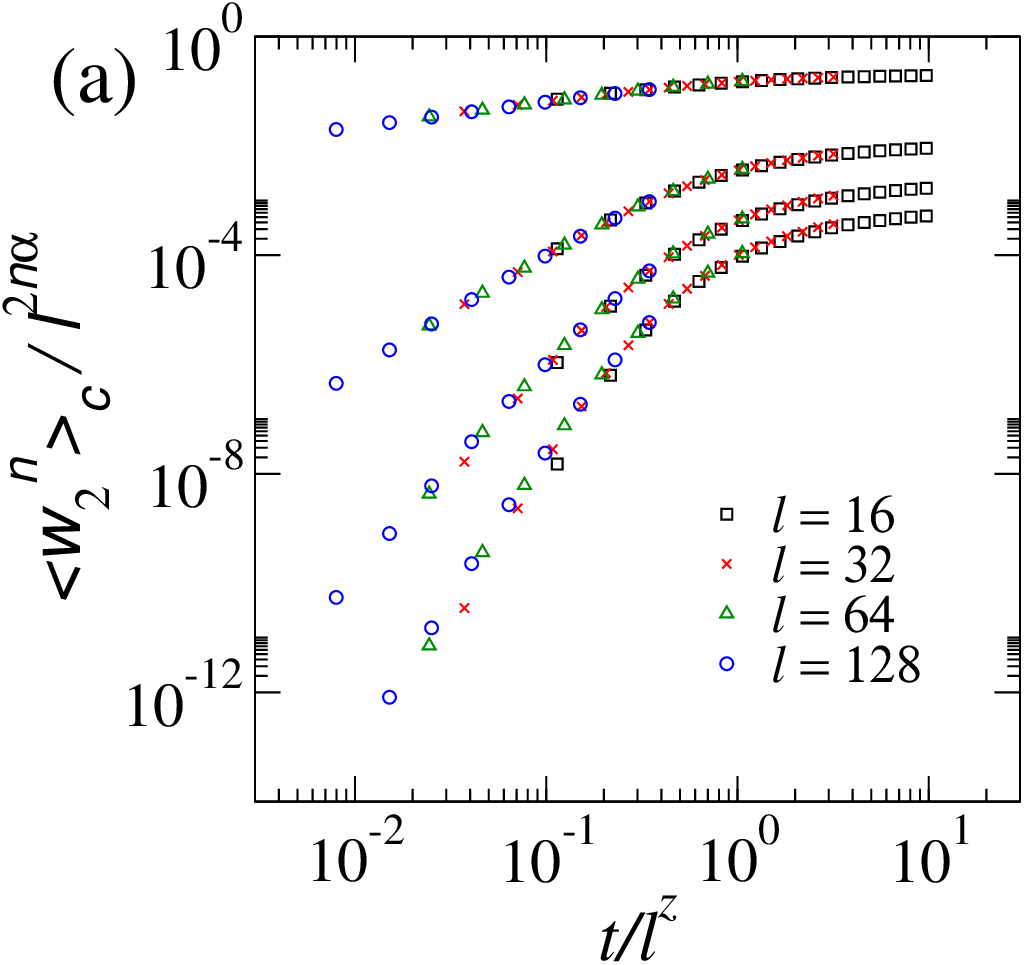}
\includegraphics[height=4.cm]{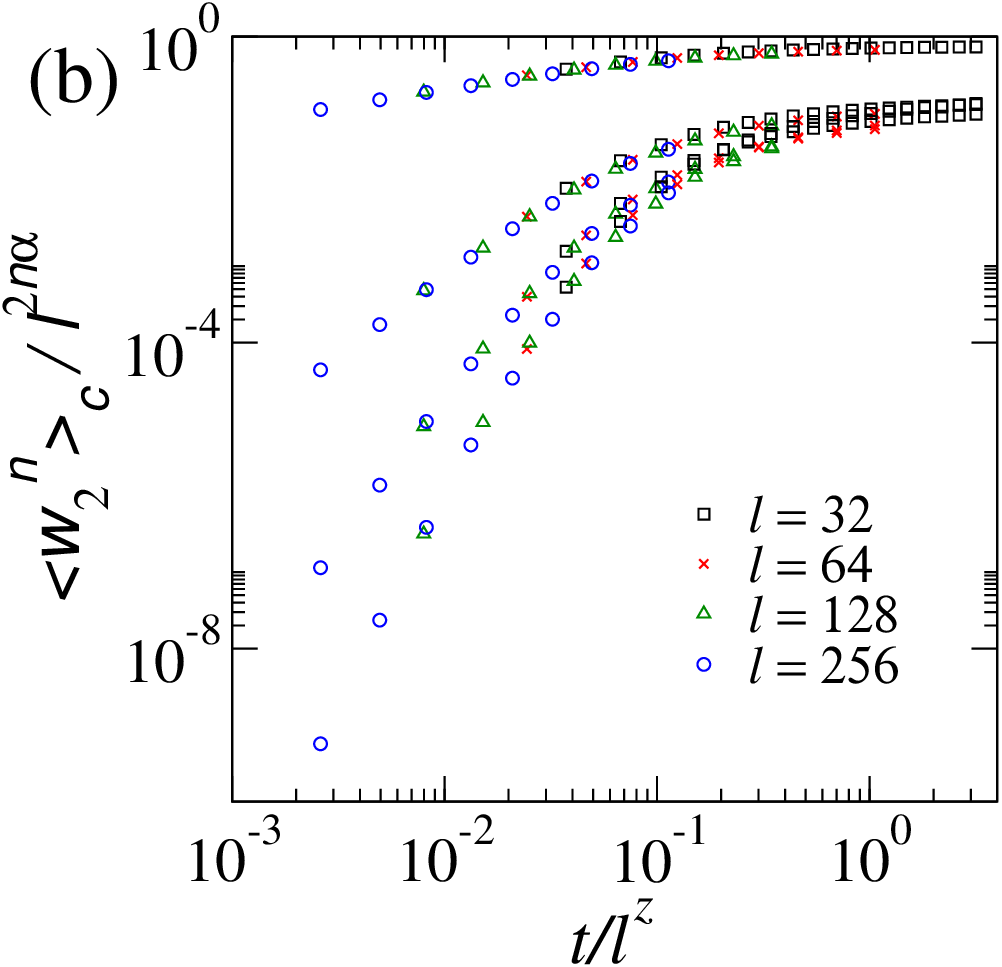}
\caption{(Color online) First four rescaled SLRDs cumulants for the (a) RSOS and (b) Etching models in $d=2+1$. From top to bottom, curves for $n=1$, 2, 3 and 4 are shown. The exponents $\alpha=0.3869$ \cite{parisi15} and $z=2-\alpha$ were used here.}
\label{fig2}
\end{figure}

From the FV scaling, we might expect, for the $n$th cumulant of SLRDs, that
\begin{equation}
 \left\langle w_2^{n} \right\rangle_c \sim l^{2 n \alpha} f_n\left( t/l^{z}\right),
 \label{eqFVgeral}
\end{equation}
as is, indeed, confirmed in the insets of Figs. \ref{fig1}a-b and Figs. \ref{fig1}c-d by the nice collapse of the rescaled cumulants for a given model, in $d=1+1$. Moreover, this scaling also holds in $d=2+1$, as shown in Figs. \ref{fig2}a and \ref{fig2}b for the RSOS and Etching models, respectively. In Fig. \ref{fig1}b, we see that $f_2(x) \approx const.$ for $t \gg l^z$, as expected. However, $\left\langle w_2^2 \right\rangle_c$ depends on both $t$ and $l$ at short times. In general, we have found $\left\langle w_2^n \right\rangle_c \sim l^{(1-n)d_s} t^{\gamma_n}$, with the exponent $\gamma_n$ depending on the exponents $\beta$ and $\alpha$ (or $z$). Thus, $f_n(x) \sim x^{\gamma_n}$, rather than the simple behavior $f_n(x) \sim x^{2 n \beta}$, which we could naively expect.

Actually, this $l$-dependence and non-trivial temporal scaling is not limited to KPZ systems, but a general feature of the cumulants of SLRDs, and also of (global) WDs, when $\xi \ll l$ (or $L$). For instance, a similar result can be proved for the EW class in $d=1+1$, for the \textit{global} WD calculated by Antal and R\'acz \cite{racz4} as a function of time. In fact, from the generating function of the moments calculated in \cite{racz4} it is straightforward to demonstrate that, at short times ($\xi \ll L$), $\left\langle W_2^n \right\rangle_c \sim L^{(1-n)} t^{\gamma_n}$, with $\gamma_n=n-1/2$ (see appendix \ref{ApEW}). Although $W_2$ is the global roughness (for PBC), we remark that, when $\xi \ll l \leq L$, the BCs becomes irrelevant, since most columns inside a given box are uncorrelated, and, thus, the same behavior shall be found for WBC. We have confirmed this through simulations of the Family model in $d=1+1$, where $\left\langle w_2^n \right\rangle_c \sim l^{(1-n)} t^{n-1/2}$ is indeed found. 

Additional proof of the $l$-dependence in high order SLRDs' cumulants is provided for the RD model, whose WD (calculated in appendix \ref{ApRD}) has $\left\langle W_2^n \right\rangle_c \sim L^{(1-n)d_s} t^{2 n \beta}$. Since all heights are uncorrelated in this system, it is obvious that WD and SLRD are equivalent, so that $\left\langle w_2^n \right\rangle_c \sim l^{(1-n)d_s} t^{2 n \beta}$. Noteworthy, the ``expected'' temporal scaling $\left\langle w_2^n \right\rangle_c \sim t^{2 n \beta}$ is found in this case, suggesting that the correlation length (which is $\xi \sim 1$ here) is the responsible for the non-trivial $\gamma_n$ exponents in the other classes. In fact, for the $1+1$ EW class, where $\beta=1/4$ and $z=2$ \cite{barabasi}, one may write $\gamma_n = n- 2 \beta = 2 n \beta + (n-1)/z$ and, then, $\left\langle w_2^n \right\rangle_c \sim [l/\xi(t)]^{(1-n)} t^{2 n \beta}$. The results for the ($1+1$) KPZ class (Fig. \ref{fig1}) are also consistent with this behavior. Therefore, the relevant measure in the finite-size scaling is $l_R = l/\xi(t)$, which gives the effective number of uncorrelated sites within a given box, so that, in general, $\left\langle w_2^n \right\rangle_c \sim l_R^{(1-n) d_s} t^{2 n \beta}$. 

\begin{figure}[!t]
\includegraphics[height=3.9cm]{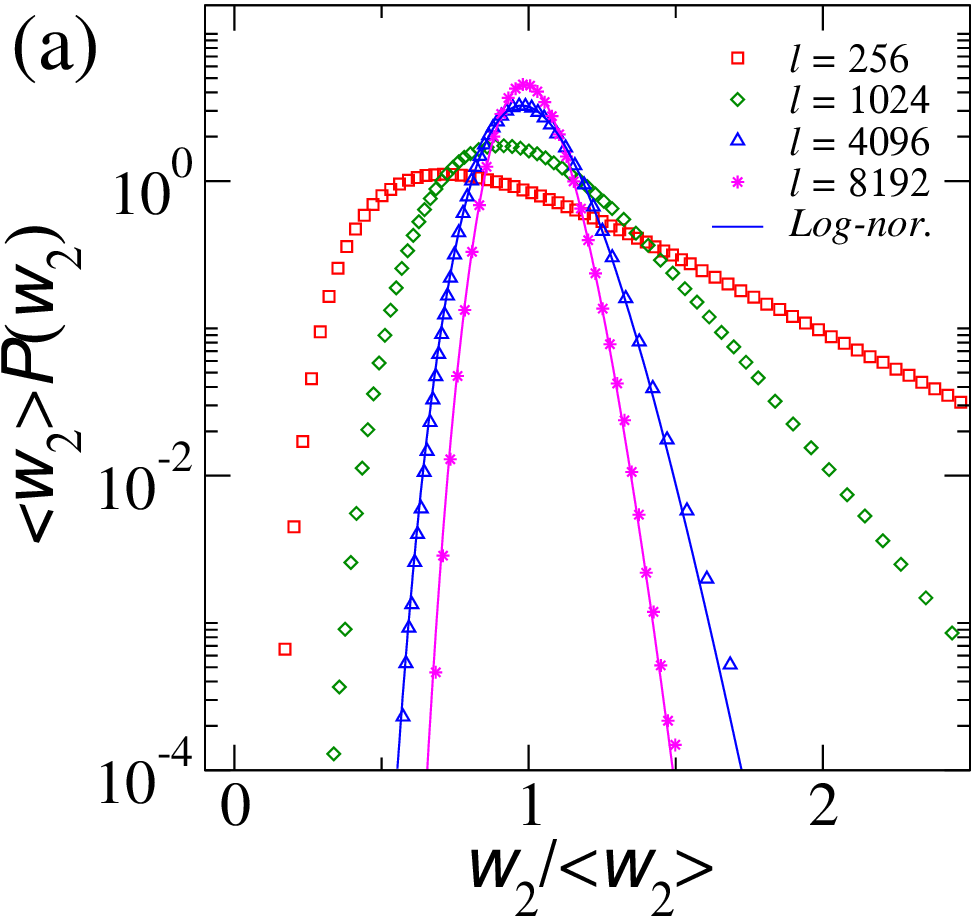}
\includegraphics[height=3.9cm]{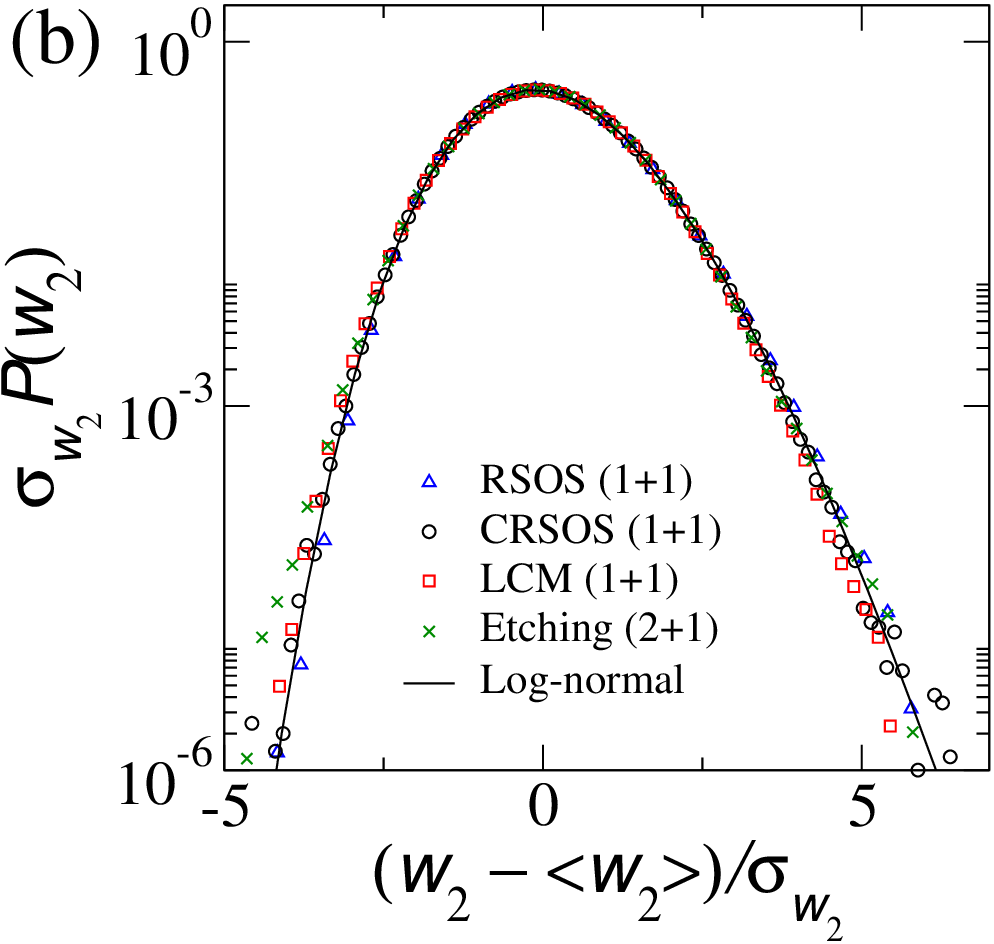}
\includegraphics[height=3.9cm]{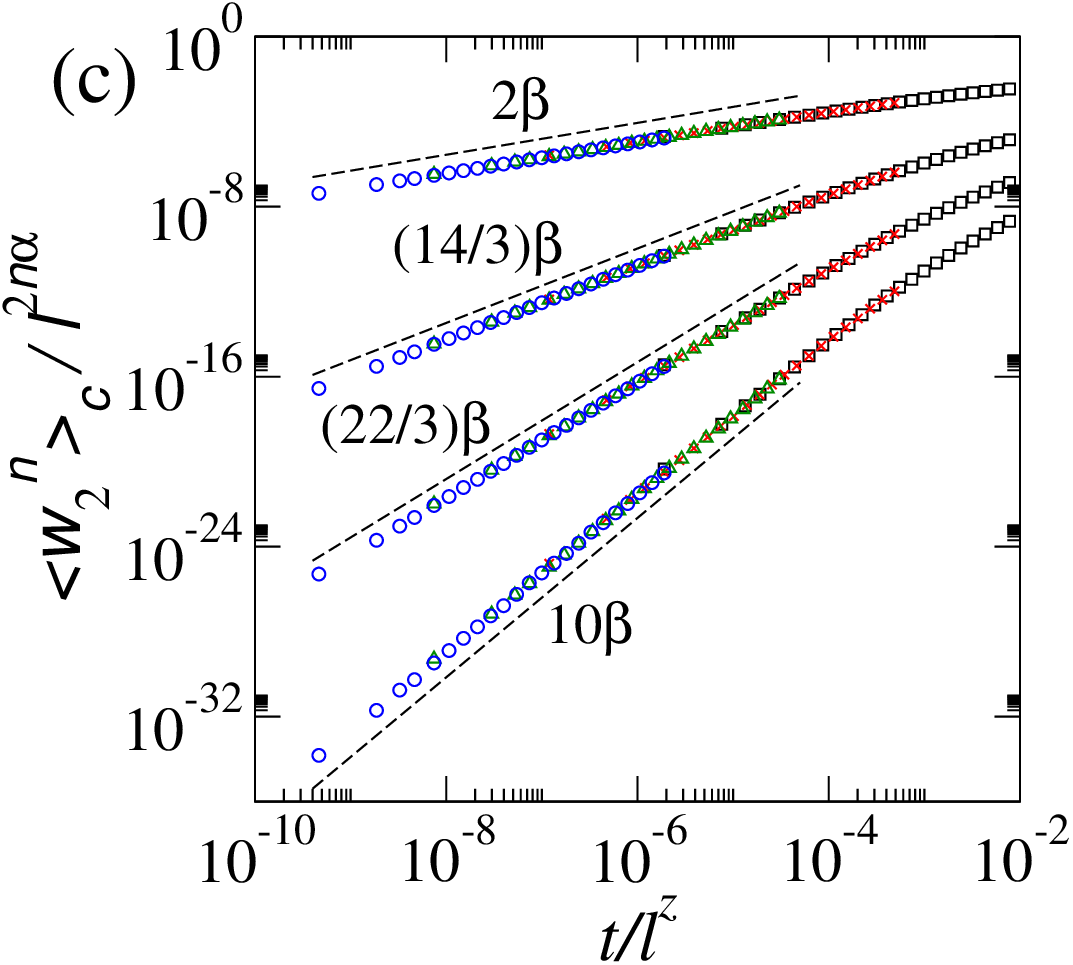}
\includegraphics[height=3.9cm]{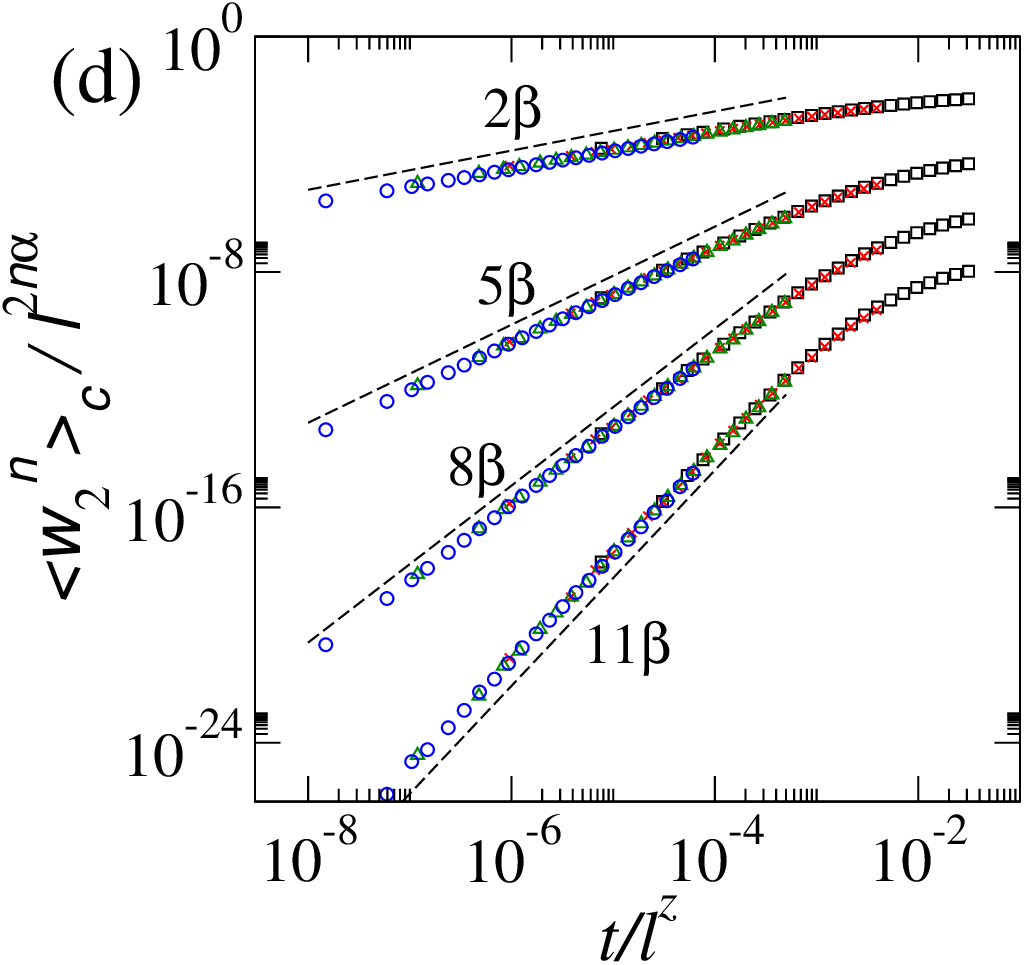}
\caption{(Color online) Rescaled SLRDs for (a) $1+1$ SS model for $t=250$ and several $l$'s and (b) for several models with $t=100$ (25) and $l=2048$ (256) in $d=1+1$ ($2+1$). First four rescaled SLRDs cumulants for the (c) LCM and (d) CRSOS models in $d=1+1$. From top to bottom, curves for $n=1$, 2, 3 and 4 are shown. The dashed lines have the indicated slopes. In (a) $\alpha=3/2$, $\beta=3/8$ and $z=4$, and in (b) $\alpha=1$, $\beta=1/3$ and $z=3$ \cite{barabasi}.}
\label{fig3}
\end{figure}

The finite-size correction in the variance, $\left\langle w_2^2 \right\rangle_c \sim 1/l_R^{d_s}$, can be simply understood noting that when $l_R^{d_s}$ random heights are summed, to obtain $\bar{h_i}$, $\bar{h_i^2}$ and, then, $w_{2,i}$, the central limit theorem states that the variance of the fluctuating variable $w_{2}$ will be of order $1/l_R^{d_s}$. So, when $l_R^{d_s} \rightarrow \infty$ (meaning that $t \rightarrow 0$ for a fixed $l$ or $l\rightarrow\infty$ for a fixed $t$) the SLRDs (and WDs) tends to a delta function, as pointed in \cite{racz4,FabioGR2}. Interestingly, Antal and R\'acz \cite{racz4} have showed that while the WD pdf of the $1+1$ EW class goes to a delta, it first approaches a log-normal distribution
\begin{equation}
P(x,t) =  \frac{1}{\sqrt{2 \pi} \sigma_x x} {\exp\left[ {- \frac{(\ln(x)-\mu)^2}{2 \sigma_x^2}}\right] },
\label{eqLogNorm}
\end{equation}
with $x=w_2/\left\langle w_2 \right\rangle$. We claim that, instead of a particularity of the $1+1$ EW class, this is a general feature of random (or almost random) interfaces. Indeed, in appendix \ref{ApRD}, we demonstrate that the SLRD/WD pdf for the RD approaches the log-normal (with $\mu=0$), when $l,L \gg 1$. Figure \ref{fig3}a shows scaled SLRDs for the $1+1$ SS (KPZ) model for a fixed $t$ and different $l$'s and, for large $l$'s (when $\xi \ll l$) they are well-fitted by Eq. \ref{eqLogNorm}. The limit $\xi \gg l$ will be discussed in Sec. \ref{secSLRDs}. The same behavior is found in all universality classes in $d=1+1$ and $2+1$, for $\xi \ll l$, as shown in Fig. \ref{fig3}b. As shown in appendix \ref{ApRD}, if the variance of the log-normal scales as  $\left\langle x^2 \right\rangle_c \sim 1/l_R^{d_s}$, the high order cumulants shall be $\left\langle x^n \right\rangle_c \sim l_R^{(1-n)d_s}$, which explains the behavior $\left\langle w_2^n \right\rangle_c \sim l_R^{(1-n) d_s} t^{2 n \beta}$. Finally, since $l_R \sim l/t^{1/z}$, in general, one has $\left\langle w_2^n \right\rangle_c \sim l^{(1-n) d_s} t^{\gamma_n}$ with
\begin{equation}
 \gamma_n = 2 n \beta + \frac{(n-1)d_s}{z} = \left[ 2 n + \frac{(n-1)d_s}{\alpha} \right] \beta.
\label{eqGamma}
\end{equation}
In fact, for the RD model, where $\alpha, z \rightarrow \infty$ \cite{barabasi}, $\gamma_n = 2 n \beta$ is recovered. For KPZ class, $\alpha=1/2$, in $d=1+1$, yields $\gamma_n = (2n-1) 2 \beta$ as, indeed, observed in Fig. \ref{fig1}. Moreover, the scaled SLRDs cumulants for the LCM model (MH class) and the CRSOS model (VLDS class), shown, respectively, in Figs. \ref{fig3}c and \ref{fig3}d, have a nice scaling with $\gamma_n^{MH}=[2(4 n -1)/3] \beta$ and $\gamma_n^{VLDS}=(3 n -1)\beta$ for several decades in time. Since, $\alpha^{MH}=3/2$ and $\alpha^{VLDS}=1$, this confirms the general validity of the scaling law (\ref{eqGamma}).

\begin{table}[!b]
\begin{center}
\begin{tabular}{cccccccccccc}
\hline\hline
model    &  &  $\gamma_1$   &  &  $\gamma_2$   &  &  $\alpha$    &  &   $\beta$     & \\
\hline
RSOS     &  &  $0.483(9)$   &  &  $2.189(5)$   &  &  $0.39(1)$   &  &   $0.242(5)$   & \\
SS       &  &  $0.48(1)$    &  &  $2.20(1)$    &  &  $0.39(1)$   &  &   $0.240(5)$   & \\
Etching  &  &  $0.47(1)$    &  &  $2.18(2)$    &  &  $0.38(2)$   &  &   $0.235(5)$   & \\ 
\hline\hline
\end{tabular}
\caption{Exponents from $\left\langle w_2^n \right\rangle_{c} \sim t^{\gamma_n}$, with $n=1$ and $n=2$, and the corresponding scaling exponents $\beta=\gamma_1/2$ and $\alpha= \gamma_1/(\gamma_2-2 \gamma_1)$ for KPZ models in $d=2+1$.}
\label{tab1}
\end{center}
\end{table}

Therefore, from estimates of the $\gamma_n$'s, it is possible to obtain the ``classical'' exponents $\alpha$, $\beta$ and $z$ from Eq. \ref{eqGamma}. For example, the values of $\gamma_n$ - calculated by estimating the effective exponents $\gamma_n^{eff}$ as the maxima of the successive slopes from the curves of $\ln \left\langle w_2^n \right\rangle_{c} \times \ln t$ and, then, extrapolating $\gamma_n^{eff}$ for large $l$'s - for KPZ models in $d=2+1$ are displayed in Tab. \ref{tab1}. The exponents $\alpha$ and $\beta$ obtained from them are in nice agreement with the best estimates known from the global roughness scaling ($\alpha = 0.3869(4)$ and $\beta \approx 0.24$) \cite{parisi15}.

We remark that the usual way to determine $\alpha$ is to grow the interface until the steady state for different substrate sizes $L$ and then use the saturation roughness ($W_{sat}$) scale: $W_{sat} \sim L^{\alpha}$. However, usually, this requires long simulational times and, thus, the best available results are limited to relatively small $L$'s, mainly in $d\geqslant 2+1$~\cite{Kelling,parisi15}, where finite-size effects can play a relevant role. More important, it is very hard to attain the steady state in experiments and, thence, the possibility of estimating $\alpha$ from a temporal scaling, as devised here, is of paramount importance.

Actually, in systems following the FV scaling, $\alpha$ can be obtained in the growth regime from the scaling of the local roughness with the box size $l$ ($\left\langle w_2 \right\rangle_c \sim l^{2 \alpha_{loc}}$, where $\alpha_{loc}$ is the local roughness exponent), since $\alpha=\alpha_{loc}$. However, this scaling usually have strong corrections and may be also featured by crossover effects due to grain/mound structures at surface \cite{tiago07Graos,*tiago11Graos2}. Even worse, several systems present anomalous scaling, so that $\alpha \neq \alpha_{loc}$ \cite{lopez99,*ramasco00}. For instance, the scaling of the LCM model (the MH class), in $d=1+1$, is anomalous and $\alpha_{loc}=1$ whereas $\alpha = 3/2$. As confirmed in Fig. \ref{fig3}c, the temporal scaling of the MH SLRDs' cumulants yields the (global) exponent $\alpha$.

Reliable estimates of the correlation length $\xi(t)$ - and, consequently, of the dynamic exponent $z$ (from $\xi \sim t^{1/z}$) - are also difficult to obtain, for example, in surfaces with multipeaked grains/mounds \cite{Almeida14,Almeida15}. So, the possibility of calculating $z$ from a simple measure such as $\left\langle w_2^n \right\rangle_c$ is very useful. Indeed, from Eq. \ref{eqGamma}, one has $z = 2/(\gamma_2 - 2 \gamma_1)$ and, then, using the values in Tab. \ref{tab1}, we find $z=1.63(3)$ (RSOS), $1.61(3)$ (SS) and $1.61(4)$ (Etching), again, in good agreement with the best known estimate ($z \approx 1.613$ \cite{parisi15}).

\subsection{LEHDs}

\begin{figure}[!t]
\includegraphics[width=4.cm]{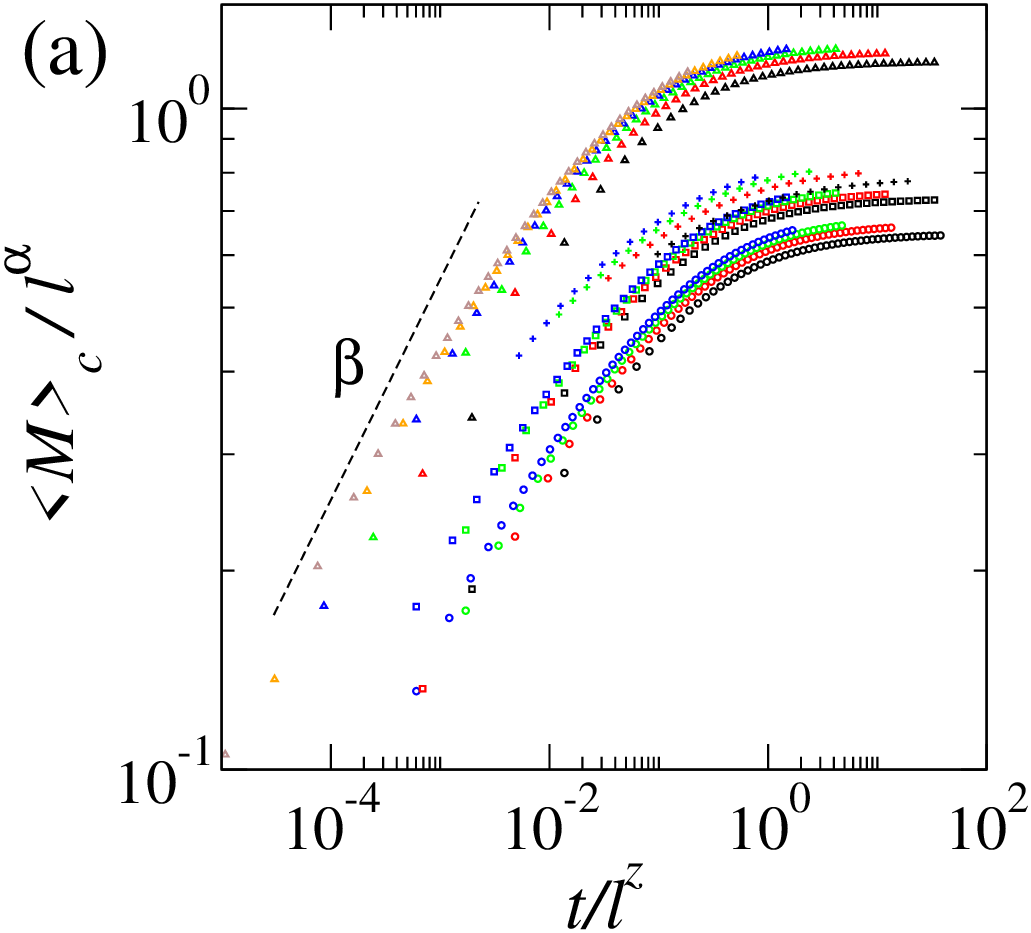}
\includegraphics[width=4.cm]{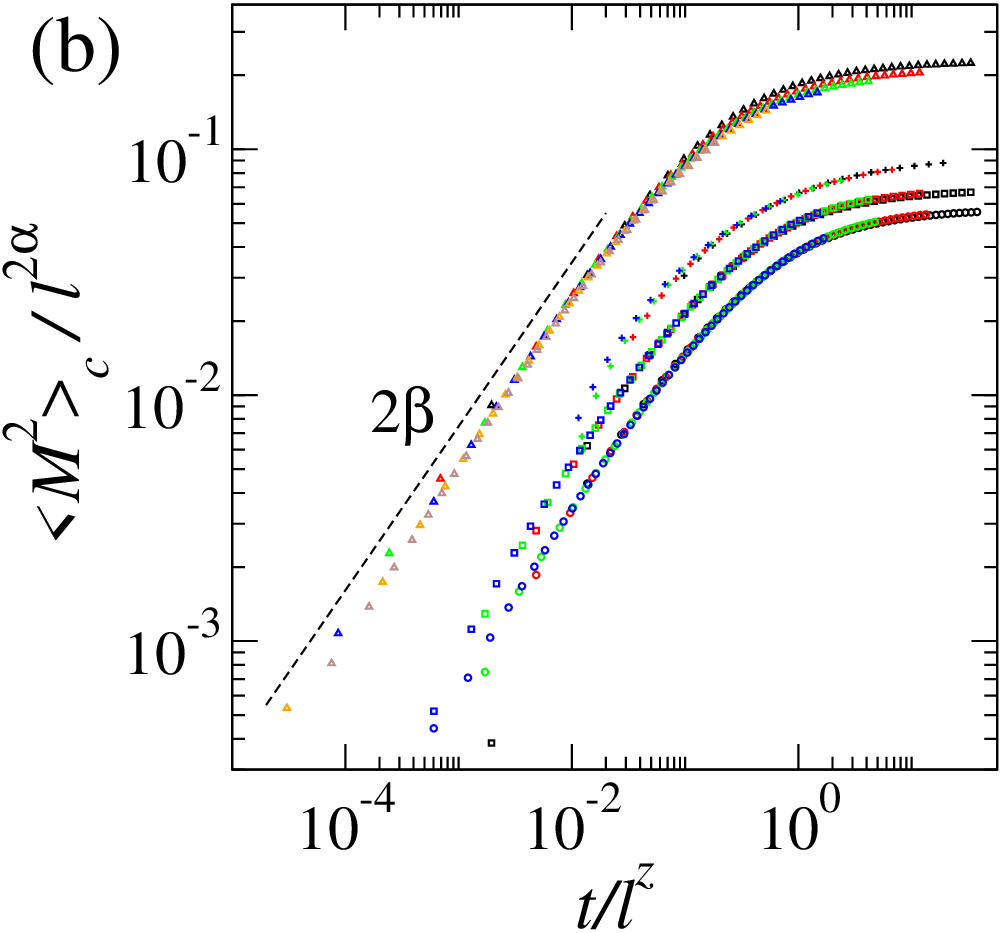}
\includegraphics[width=4.cm]{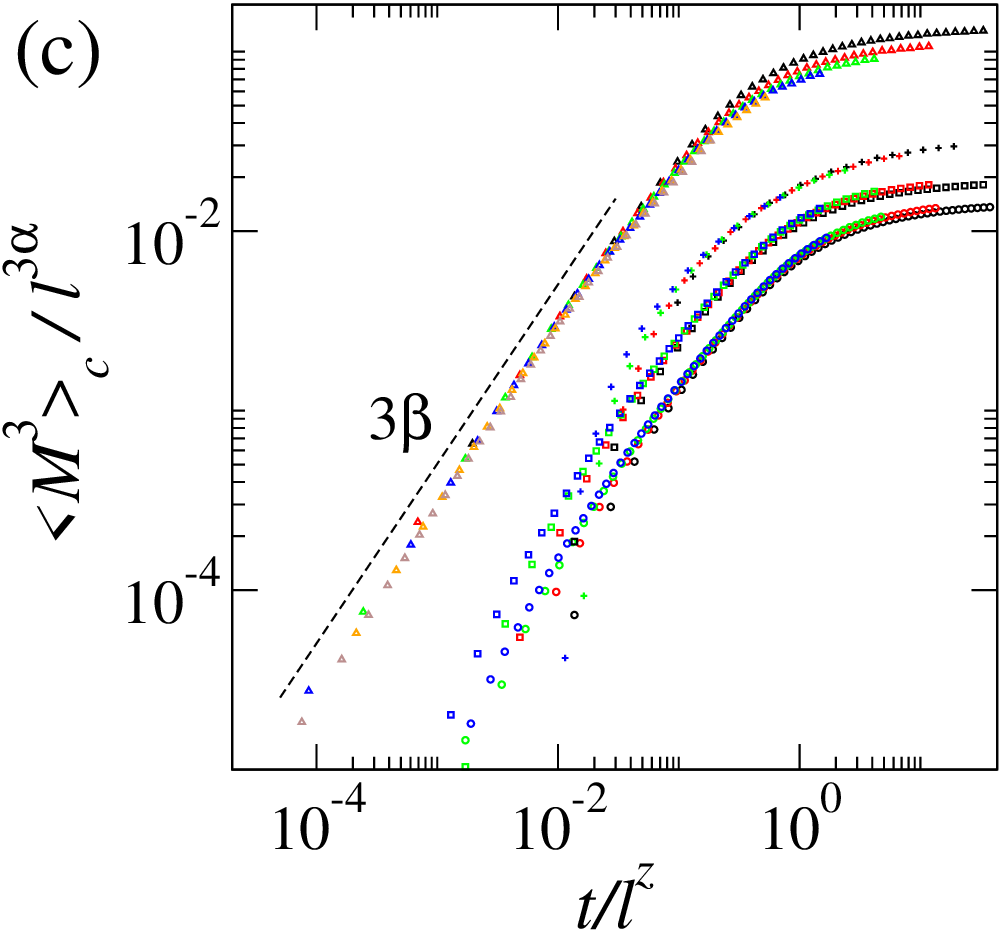}
\includegraphics[width=4.cm]{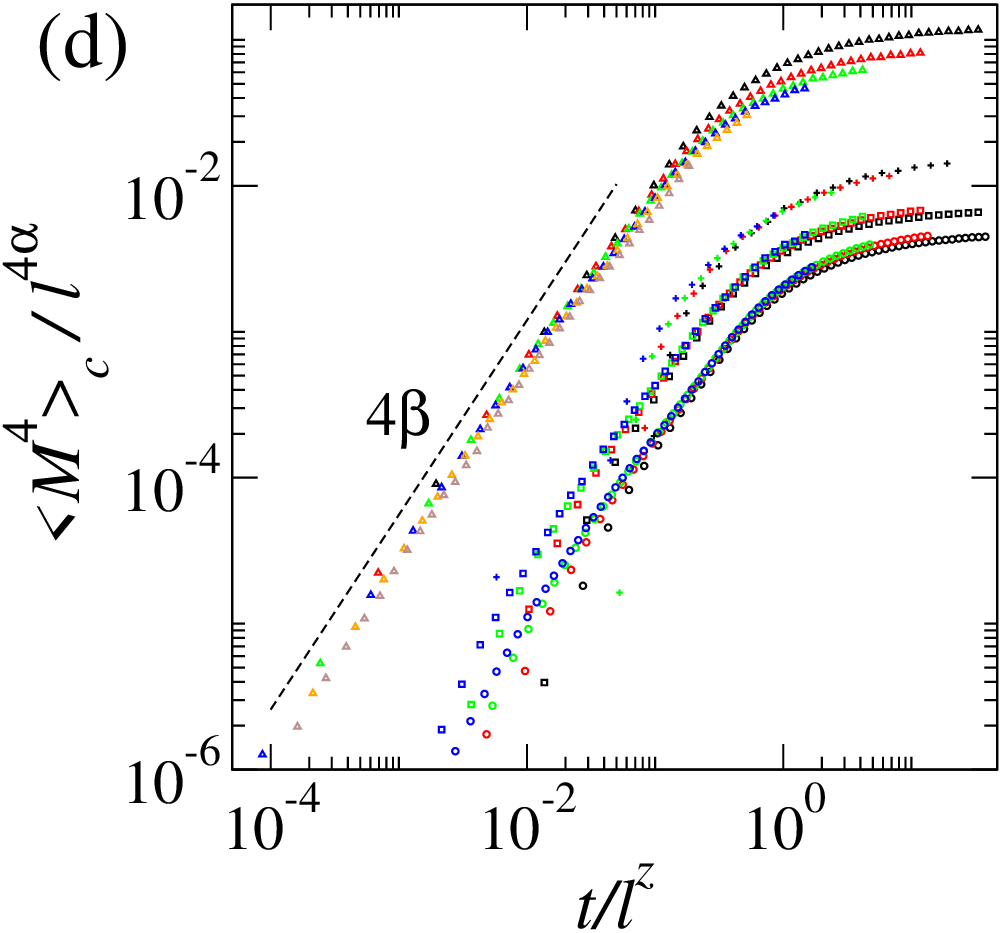}
\caption{(Color online) (a) First, (b) second, (c) third and (d) forth rescaled cumulants of MAHDs, calculated in different box sizes $l$. From bottom to top, collapsed curves for RSOS, SS, Eden and Etching models in $d=1+1$ are shown. The dashed lines have the indicated slopes, where $\beta=1/3$.}
\label{fig4}
\end{figure}

At the PBC steady state, the mean (global) maximal relative height $\left\langle M_{sat} \right\rangle_c $ scales as the surface roughness $\left\langle M_{sat} \right\rangle_c \sim \left\langle W_{sat} \right\rangle \sim L^{\alpha}$ \cite{raychaudhuri01}. The only known exception is the EW class in $d=2+1$, where $\left\langle W_{sat} \right\rangle \sim \sqrt{\ln{L}}$ and $\left\langle M_{sat} \right\rangle_c \sim \ln{L}$ \cite{lee05,tiago08Ext}. Anyway, in most cases, this suggests for the cumulants of the local MAHDs that
\begin{equation}
 \left\langle M^{n} \right\rangle_c \sim l^{n \alpha} g_n\left( t/l^{z}\right),
 \label{eqFVgeralMAX}
\end{equation}
where the scaling function $g_n(x)$ would scale as $g_n(x) \sim const.$ for $x \gg 1$ and $g_n(x) \sim x^{\delta_n}$ for $x \ll 1$. This scaling is confirmed in Figs. \ref{fig4}a-\ref{fig4}d by the good collapse of the high order ($n>1$) rescaled cumulants of the MAHDs for the KPZ models in $d=1+1$. However, for the mean $n=1$ (Fig. \ref{fig4}a) reasonable collapses are observed only for the larger $l$'s. Indeed, Raychaudhuri et al. \cite{raychaudhuri01} demonstrated that the (global) maximal relative height increases in time as $\left\langle M_g \right\rangle  \sim t^{\beta} \left[ C + \ln{L} - \left( \beta/\alpha \right) \ln{t} \right]^{1/a} $, where $C$ is a constant and the exponent $a$ is associated to the decay of the right tail either of the extremal height distribution or the velocity distribution. Therefore, the origin of the deviation in $\left\langle M \right\rangle_c$ (Fig. \ref{fig4}a) is certainly these logarithmic enhancements. It is impressive that such logarithmic seem not be present in the high order cumulants.

As a consequence of the finite-time and -size effects, the scaling of $\left\langle M \right\rangle_c$ in time is not so clear, but is initially consistent with $\left\langle M \right\rangle_c \sim t^{\beta}$ (see Fig. \ref{fig4}a). On the other hand, for the higher order cumulants (Figs. \ref{fig4}b-\ref{fig4}d), we find good power laws $\left\langle M^{n} \right\rangle_c \sim t^{n \beta}$. Thus, $g_n(x) \sim x^{n\beta}$ for $x \ll 1$, meaning that (asymptotically) $\left\langle M^{n} \right\rangle_c$ does not depend on $l$, for $t \ll l^{z}$, in contrast with $\left\langle w_2^{n} \right\rangle_c$. This happens because fluctuations in $M_i$ is dominated by $h_{max,i}$, which is not an average and, so, does not follow the central limit theorem. Therefore, it is not possible to determine $\alpha$ (and $z$) directly from $\left\langle M^{n} \right\rangle_c$ \textit{vs.} $t$, but alternative measures of $\beta$ can be found.

For the RSOS, SS and Eden models, Fig. \ref{fig4} shows that $\left\langle M^{n} \right\rangle_c$ deviate from the scaling at early times. This is due to the very smooth surfaces produced by these models at short times, which are almost flat inside a box ($M_i \lesssim 1$). In contrast, even for small $t$, the interfaces of the Etching model have a considerable roughness and, consequently, a well-behaved $\left\langle M^{n} \right\rangle_c$ scaling.

\begin{figure}[!t]
\includegraphics[width=8.1cm]{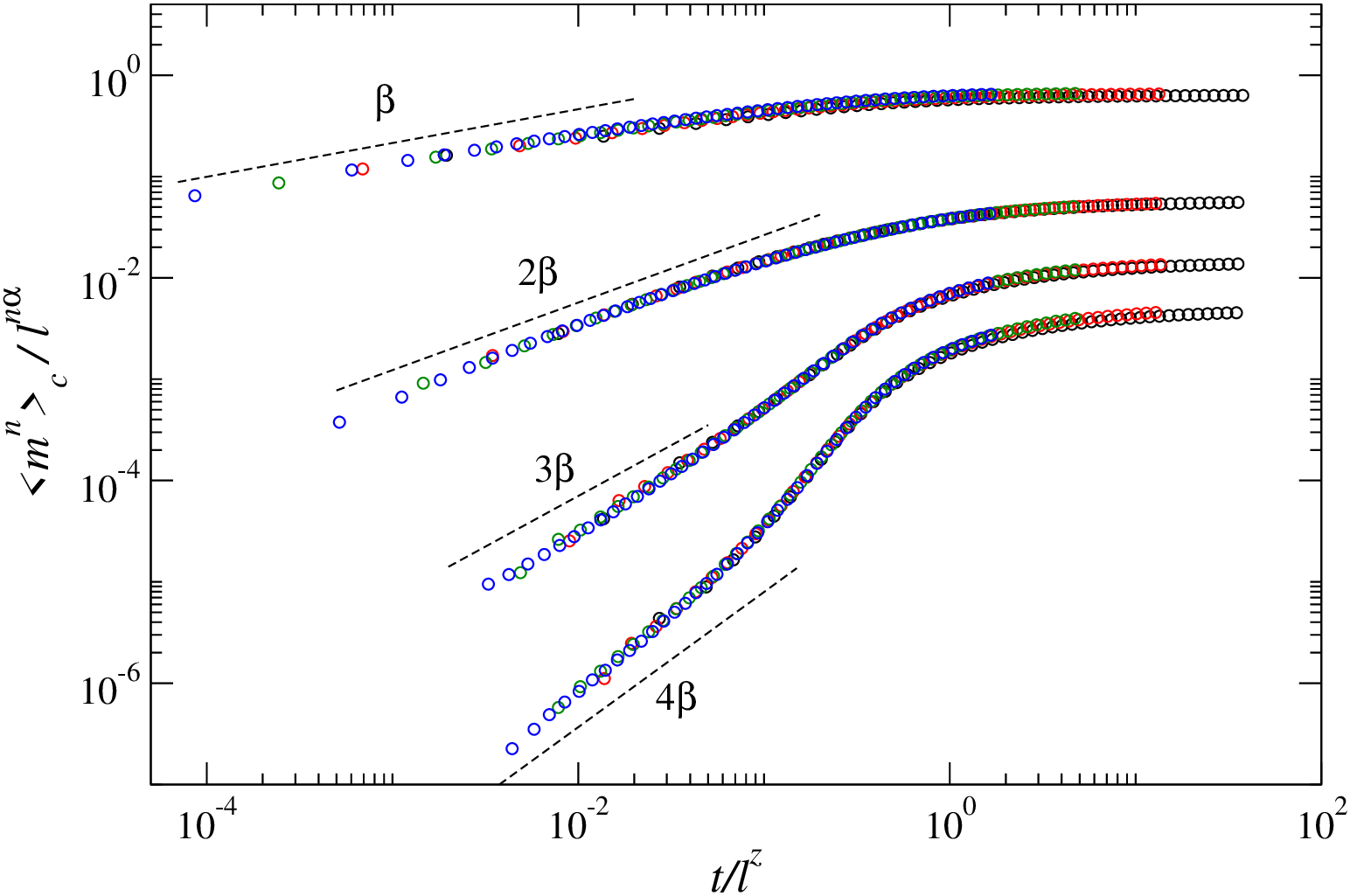}
\caption{(Color online) First four rescaled cumulants of MIHDs for the RSOS model in $d=1+1$. From top to bottom, curves for $n=1$, 2, 3 and 4 are shown. The dashed lines have the indicated slopes, with $\beta=1/3$.}
\label{fig5}
\end{figure}

The cumulants of the local MIHDs, $\left\langle m^{n} \right\rangle_c$, follows a scale similar to Eq. \ref{eqFVgeralMAX}, as shown in Fig. \ref{fig5} for the RSOS model. However, the scaling functions $g_n^{min}(x)$ for $n=3$ and $4$ have a crossover at $t \ll l^{z}$, leading to deviations of the scaling $g_n^{min}(x) \sim x^{n \beta}$. Similar results were found for the other models in $d=1+1$.

In two dimensions, the cumulants of MAHDs and MIHDs present stronger corrections than in $d=1+1$ and rescaling them according to Eq. \ref{eqFVgeralMAX} does not lead to good data collapse (not shown). So, in general, we may conclude that while the scaling of cumulants of the SLRDs is a powerful method to obtain the scaling exponents, the same does not happen with the one of the LEHDs.

\section{Squared Local roughness distributions}
\label{secSLRDs}


\begin{figure}[!t]
\includegraphics[width=8.1cm]{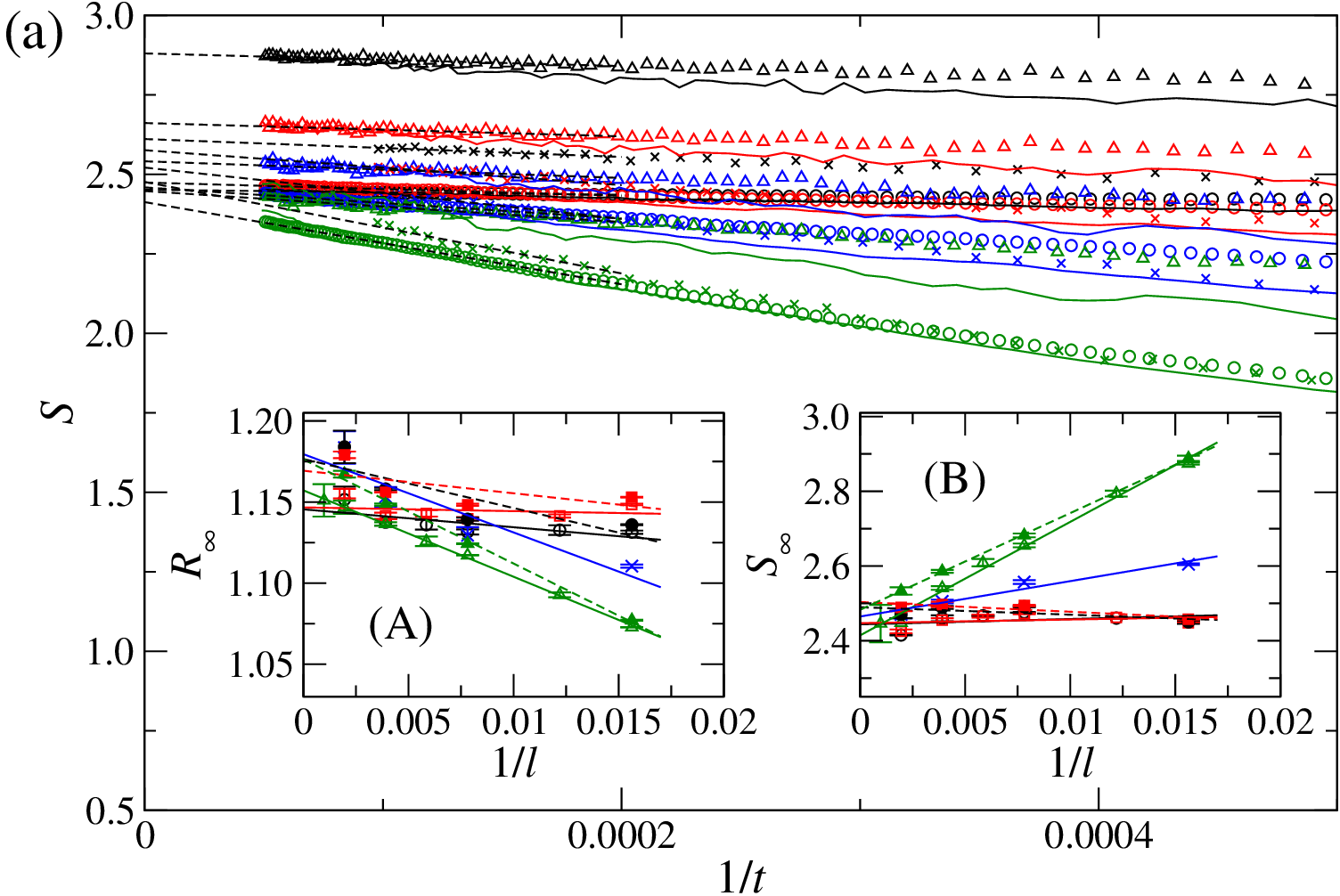}
\includegraphics[width=8.1cm]{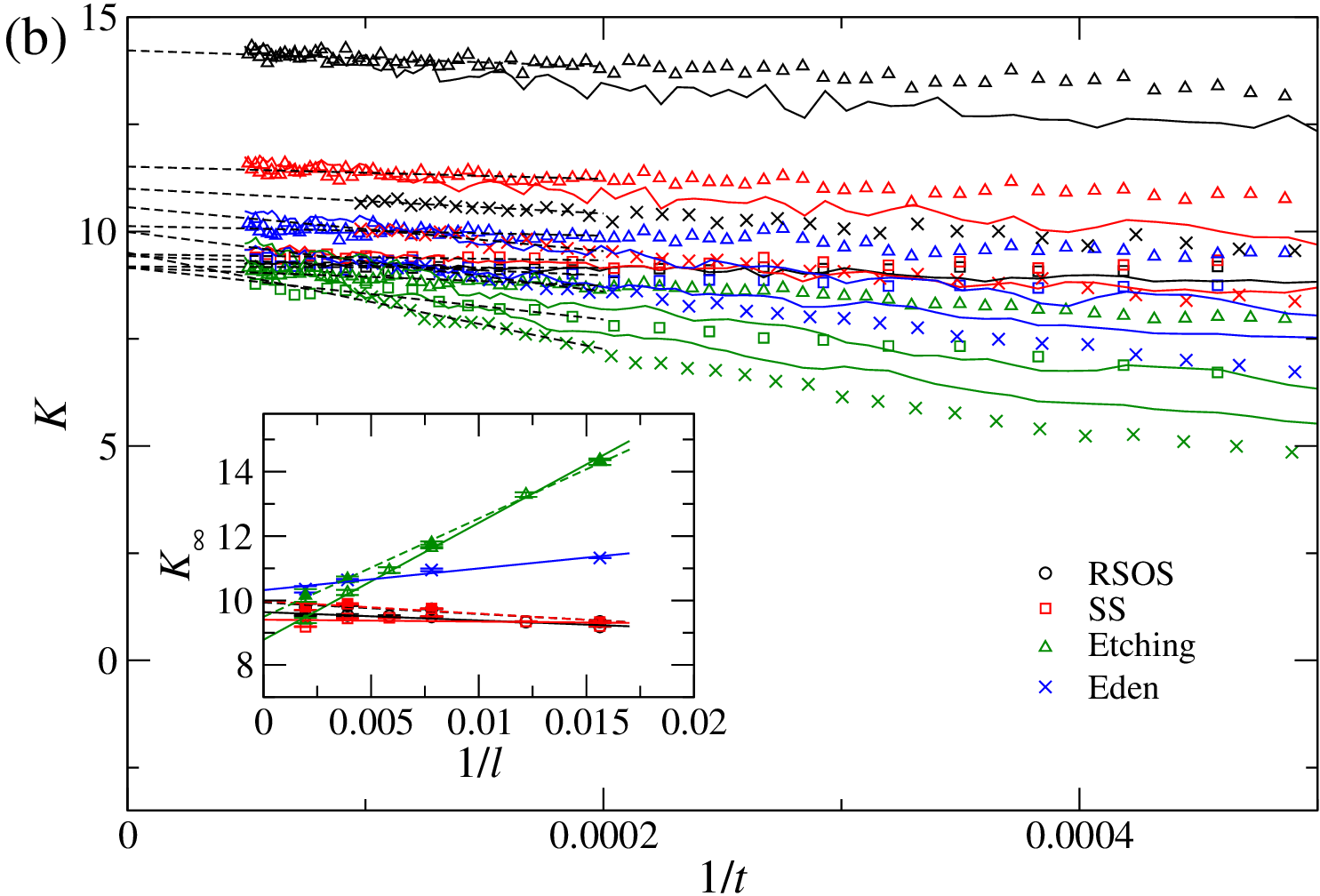}
\caption{(Color online) (a) Skewness and (b) kurtosis as functions of $1/t$ for all KPZ models in $d=1+1$, calculated in boxes of sizes $l=64$ (black), $128$ (red), $256$ (blue) and $512$ (green). For RSOS, SS and Etching models, symbols are for growth on fixed size substrates, whilst full lines for enlarging ones. In (a), the insets show the extrapolated (in time) ratios (A) $R_{\infty}$ and (B) $S_{\infty}$. The extrapolated kurtosis are displayed in the inset of (b). In insets, open and full symbols are data for fixed size and enlarging substrates, respectively, while full and dashed lines are linear fits of them.}
\label{fig6}
\end{figure}

Figures \ref{fig6}a and \ref{fig6}b show the extrapolation of the skewness $S$ and kurtosis $K$ of the SLRDs, for $t \rightarrow \infty$ (i. e., for the regime of $\xi \gg l$), for the KPZ models in $d=1+1$ in both fixed size (symbols) and enlarging substrates (full lines). In all cases, good linear behaviors are found if we use $1/t$ in the abscissa, for long times. The extrapolated values $S_{\infty}$, $K_{\infty}$, and also of the ratio $R_{\infty}$, are displayed in the insets of Fig. \ref{fig6}. For RSOS and SS models, they present negligible $l$-dependences, allowing us to obtain accurate estimates of those ratios, whose averages are depicted in Tab. \ref{tab2}. On the other hand, stronger finite-size effects are observed in SLRDs of Eden and (mainly) of the Etching models. In the latter, this is certainly due to a large intrinsic width $w^2_i$ dominating the roughness at short scales. In fact, following the procedures and definitions in Ref. \cite{Alves14BD}, we estimate here $\left\langle w^2_i \right\rangle  = 2.21(2)$ for the Etching model (in $d=1+1$). However, contrarily to its original version, for the Eden model, we do not find a correction consistent with an intrinsic width, so, it has other origin, possibly the existence of some remaining anisotropy at short scales. Anyway, for large $l$'s, we may observe those ratios converging towards the ones for SS/RSOS models. As noted by Halpin-Healy and Takeuchi~\cite{HHTake2015}, the SLRD by Antal \textit{et al.} \cite{racz5} - calculated for steady state EW interfaces with WBC - should be also the pdf of $1+1$ SLRDs of KPZ class. Indeed, this pdf has ratios $R=1.12$, $S=2.55$ and $K=10.27$ that, although do not agree with our estimates within the error bars, are very close to them. A similar slight difference has been reported for the Euler integration of the KPZ equation~\cite{HHTake2015}.

\begin{figure}[!t]
\includegraphics[width=4.2cm]{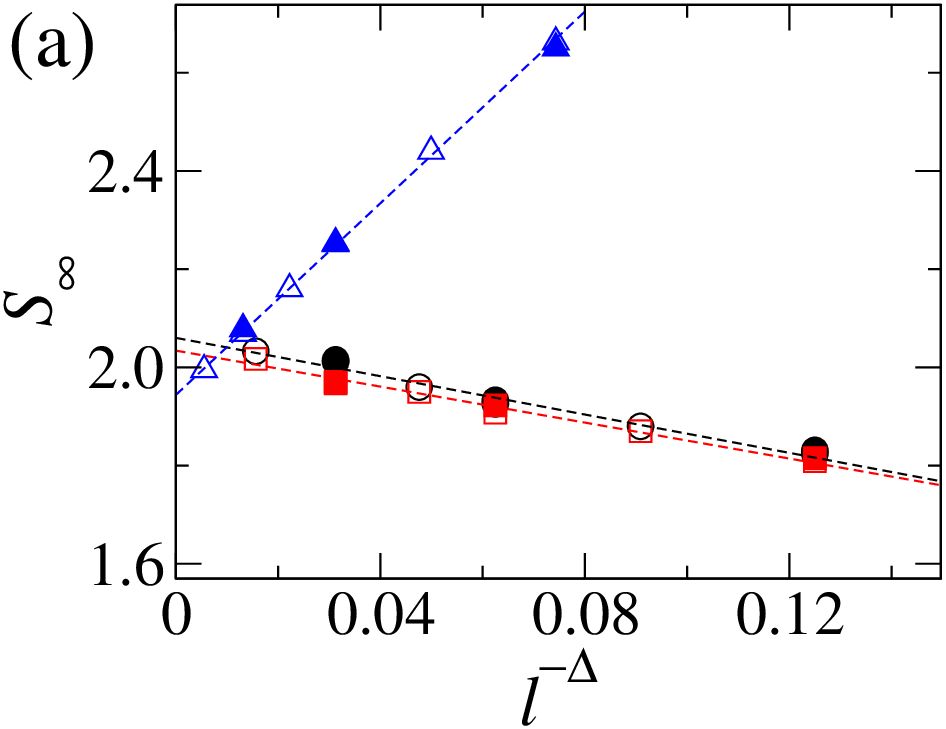}
\includegraphics[width=4.2cm]{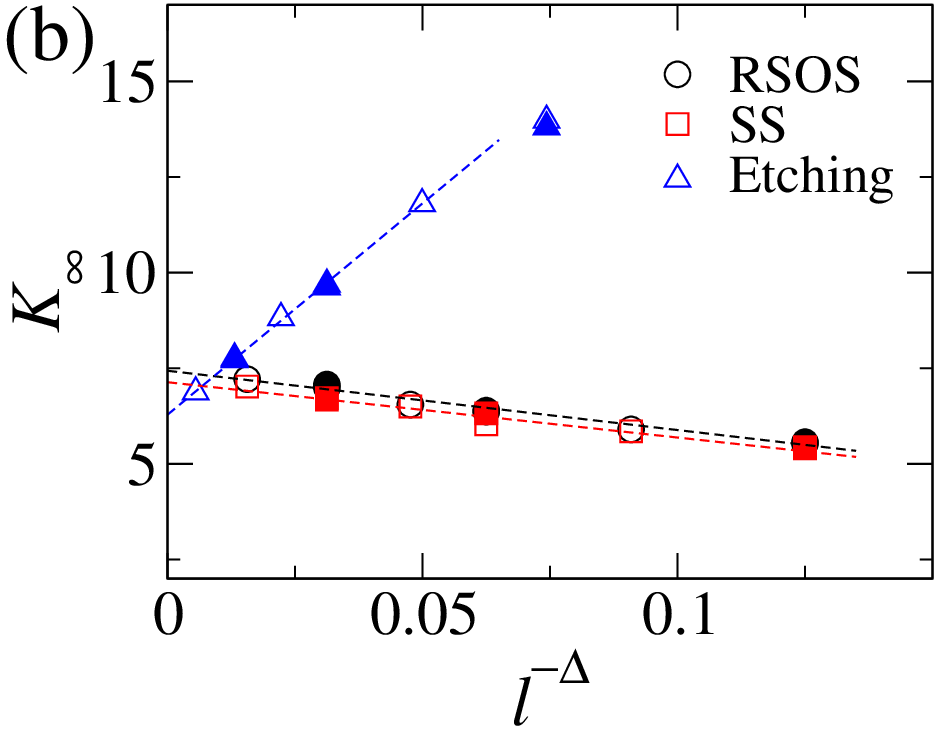}
\caption{(Color online) Extrapolated ($t \rightarrow \infty$) (a) skewness and (b) kurtosis of SLRDs as functions of $1/l^{\Delta}$, for KPZ models grown on fixed size (open) and enlarging substrates (full symbols) in $d=2+1$. The dashed lines are linear fits of the data. $\Delta=1$ for RSOS and SS and $\Delta=1.25$ for Etching model.}
\label{fig7}
\end{figure}

In $d=2+1$, we find $R$, $S$ and $K$ extrapolating nicely, again, as $1/t$. The obtained $S_{\infty}$ and $K_{\infty}$ are displayed in Figs. \ref{fig7}a and \ref{fig7}b, respectively, where some finite-size effects are observed even for the RSOS and SS models, and more severe corrections in the Etching one (in this case we estimate $\left\langle w^2_i \right\rangle  = 2.7(1)$). Notwithstanding, again, as size increases these ratios converges to similar asymptotic places. The estimates for RSOS and SS models are depicted in Tab. \ref{tab2}, where we see that $S$ and $K$ as a nice agreement with the estimates from the Euler integration of KPZ equation \cite{healy14} ($S=2.03$ and $K=7.11$). Moreover, the $R$ value is a bit larger than the estimate $1/R = 0.53(2)$ in \cite{FabioGR2}. For the Etching model, despite the stronger corrections, we find $R \approx 2.08$, $S \approx 1.95$ and $K \approx 6.4$. Whilst $R$ agrees with other models, the difference in $S$ and $K$ from the bottom limits are, respectively, $2.5$\% and $8.5$\%.

\begin{table}[!b]
\begin{center}
\begin{tabular}{cccccccc}
\hline\hline
dimension    &  &     $R$   &  &  $S$   &  &  $K$   & \\
\hline
$1+1$        &  &  $1.15(2)$   &  &  $2.44(4)$   &  &  $9.5(4)$  & \\
$2+1$        &  &  $2.05(5)$   &  &  $2.04(4)$   &  &  $7.3(3)$  & \\
\hline\hline
\end{tabular}
\caption{Cumulant ratios for KPZ SLRDs.}
\label{tab2}
\end{center}
\end{table}

These results show that our extrapolation procedure (first in time, (A) to guarantee that $\xi \gg l$ and, then, for large $l$'s, (B) to overcome finite-size effects) is a reliable way to access the universality of local distributions. We recall that in recent works \cite{Almeida14,healy14,HHTake2015,Iuri15,FabioGR2} this was achieved by performing simulations for very long growth times, to fulfill the requisite (A), and for several box sizes, in order (to try) to determine a plateau - a $l$-independent value of $R$, $S$ and $K$. Temporal extrapolations of the maxima (when they exist) of these plateaus can be also worthy \cite{FabioGR2}. However, to observe clear (wide) plateaus in systems with strong finite-size corrections, huge growth times can be necessary. This was, indeed, observed for the ($2+1$) Etching model in \cite{FabioGR2}. A similar problem shall happen in systems with large $z$, whose $\xi$'s increase very slowly in time. Moreover, in experiments, the limited number of samples and/or small $\xi$'s might prevent the observation of reliable plateaus.

On the other hand, within our framework, the temporal extrapolation provides good estimates of the cumulant ratios (for $\xi \gg l$) from data for (relatively) short deposition times. For comparison, Aar\~ao Reis \cite{FabioGR2} has obtained $2.0 \leq S \leq 2.2$ from simulations of the RSOS model, in $d=2+1$, for times up to $t=13000$, while we find here $S=2.04(4)$ from data for $t=1000$. Obviously, the temporal extrapolations cannot be done for arbitrarily short times, but we observe that working with $t$'s so that $\xi(t) \sim l$ is enough, which is much easier to work than $\xi \gg l$. For instance, recalling that $\xi = (\sqrt{A} |\lambda| t)^{1/z}$ \cite{KrugPRA92} and using the values of $A$ and $\lambda$ in \cite{Ismael14}, for $t=1000$ (in $d=2+1$), one has $\xi_{SS} \approx 52$ and good extrapolations were possible for $l \lesssim 64$. In $d=1+1$, $\xi_{SS} \approx 737$ for $t=20000$ and we have worked with $l \leq 512$.

\begin{figure}[!t]
\includegraphics[width=8.1cm]{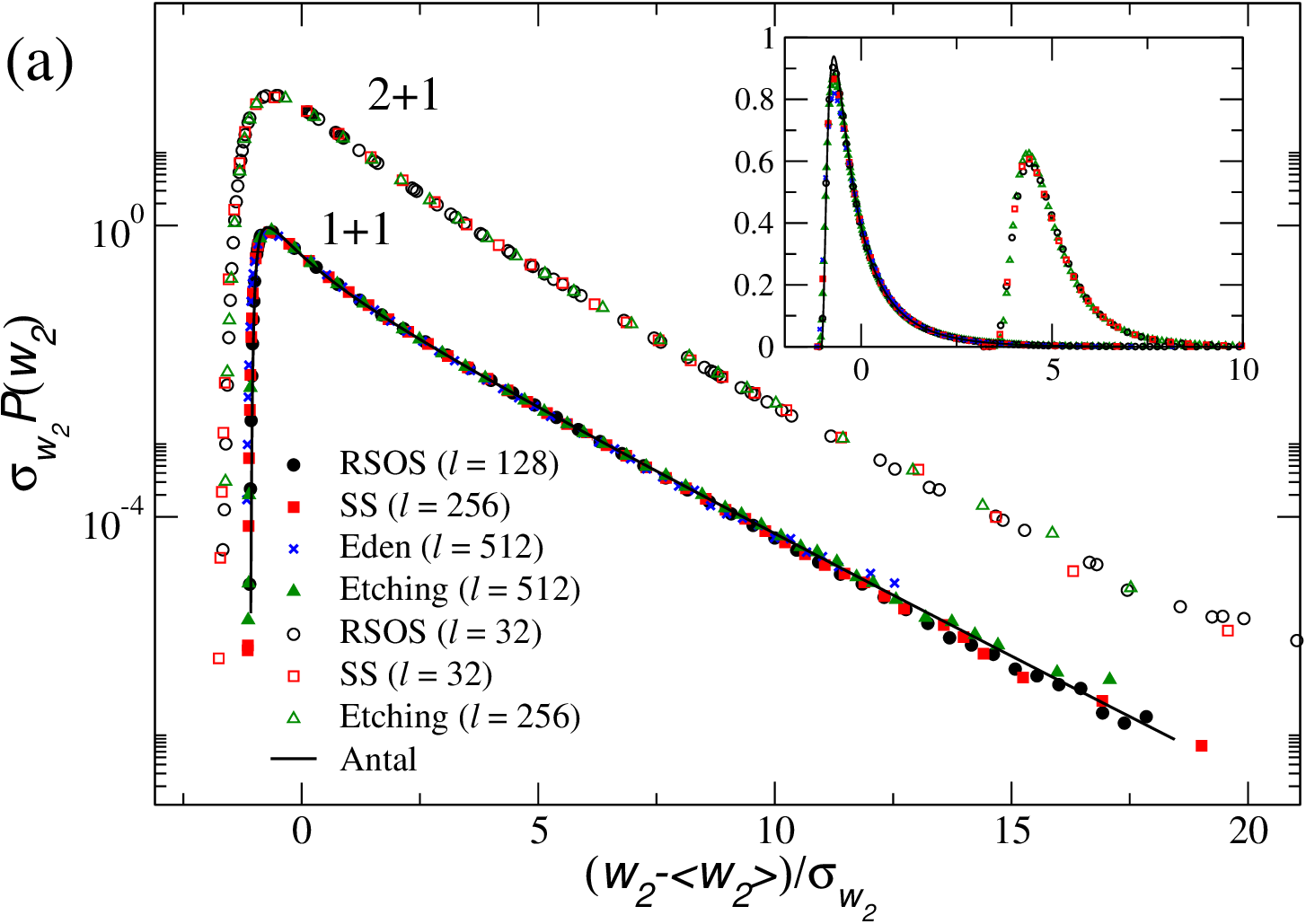}
\caption{(Color online) Rescaled SLRDs for KPZ models in $d=1+1$ (bottom) and $d=2+1$ (top). Data for $d=2+1$ was shifted up two decades. Inset shows the same SLRDs in linear scaling, with data for $d=2+1$ centered in 5. Data for fixed size (RSOS and Etching) and enlarging substrates (SS) are shown.}
\label{fig8}
\end{figure}

Another interesting finding here is that the asymptotic KPZ SLRDs are the same for flat (fixed size) and curved (enlarging substrates) geometries. Indeed, in Fig. \ref{fig6}, we see that for long times $S$ (and also $K$), for the same model and $l$, tends to become equal regardless the substrate enlarges or not. This leads to temporal extrapolations $S_{\infty}$ and $K_{\infty}$ very similar for both subclasses. In $d=2+1$, this is even more evident (see the superimposed data in Fig. \ref{fig7}). Moreover, the asymptotic cumulant ratios for the (really curved) Eden surfaces agree quite well with the ones for other (flat surface) models, in $d=1+1$.

The rescaled SLRDs in $d=1+1$ and $d=2+1$ are shown in Fig. \ref{fig8}. The nice collapse of the distributions for different models, boxes sizes and geometries gives a final confirmation of their universality in both dimensions. We remark that SRLDs for Eden and Etching models for small $l$'s (not shown) do not present a good collapse, as expected from the corrections observed in their cumulant ratios. The Antal \textit{et al.} \cite{racz5} pdf is also shown in Fig. \ref{fig8} and presents an excellent agreement with the $1+1$ SLRDs. Thus, despite the small differences in their cumulant ratios, our results confirm the claim of Halpin-Healy and Takeuchi \cite{HHTake2015} that the steady state EW distribution also plays the role in KPZ growth regime, when WBC is considered. Concerning the right tail of the SLRDs, we find evidences of an exponential (in $d=1+1$) and stretched exponential (in $d=2+1$) decay, as also suggested in \cite{HHTake2015} and \cite{FabioGR,healy14}, respectively.

\section{Extremal relative height distributions}
\label{secLEHDs}

We start this section remarking that the MAHDs and MIHDs are directly related to the decay of the right and left tails, respectively, of the HDs. Since RSOS and SS models have HDs with $S <0$, their left (right) tails are equivalent to the right (left) ones of Eden and Etching models (whose HDs have $S>0$). Therefore, we will compare MIHDs of the former models with the MAHDs of the last ones, and vice-versa.

\begin{figure}[!t]
\includegraphics[width=4.cm]{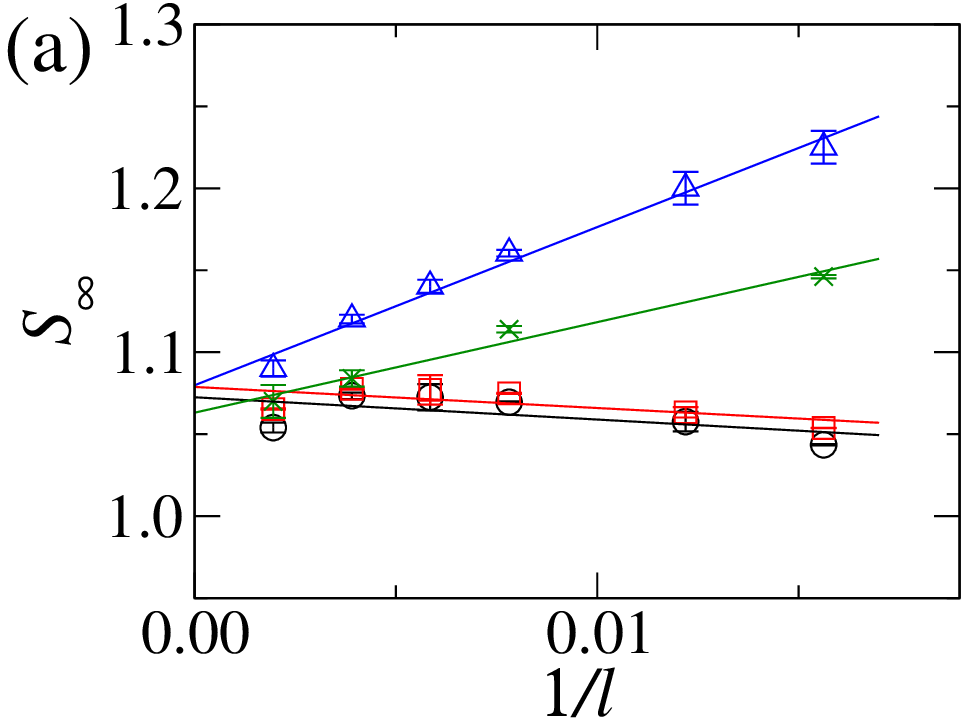}
\includegraphics[width=4.cm]{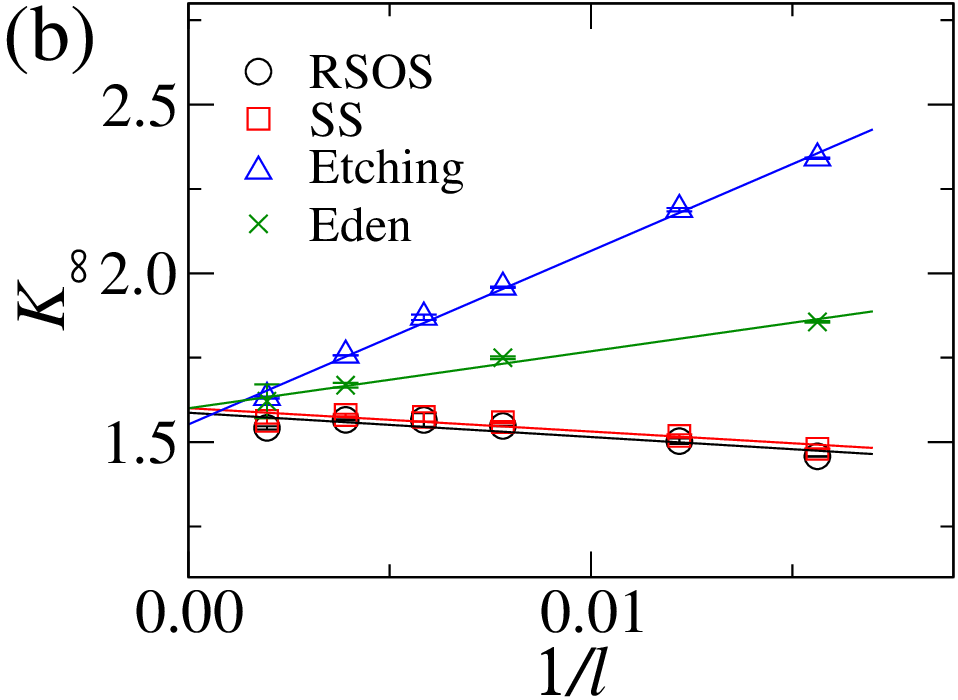}
\includegraphics[width=4.cm]{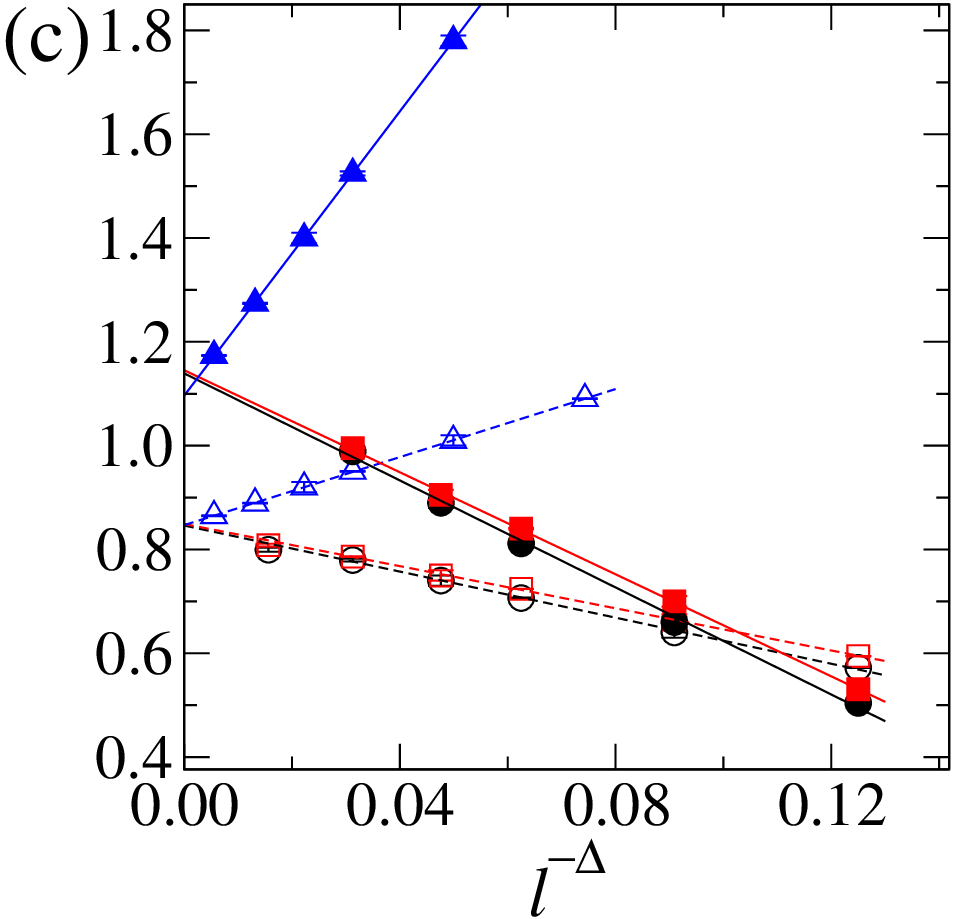}
\includegraphics[width=4.cm]{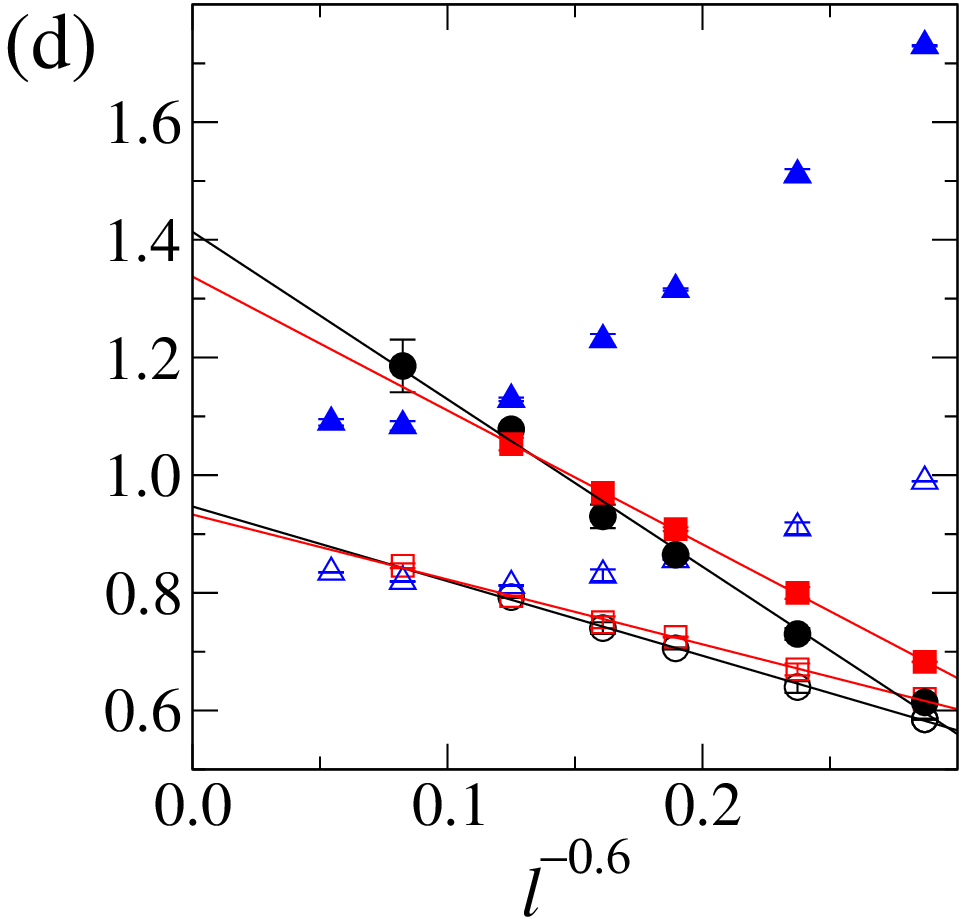}
\caption{(Color online) Extrapolated (a) skewness and (b) kurtosis for LEHDs of KPZ models in $d=1+1$. For $d=2+1$, the extrapolated skewness (open) and kurtosis (full symbols) are shown in (c) for MAHDs and (d) for MIHDs. In (c), $\Delta=1$ for RSOS/SS and $\Delta=1.25$ for Etching model.}
\label{fig9}
\end{figure}

Performing an analysis of cumulant ratios of local MAHDs and MIHDs similar to the previous section, we find that they also extrapolate in time as $1/t$. The long time values obtained for $S$ and $K$ in $d=1+1$ are shown in Figs. \ref{fig9}a and \ref{fig9}b, respectively. Again, strong finite-size corrections are found for Eden and Etching models, but, after a size extrapolation, we find accurate estimates of $R$, $S$ and $K$ (depicted in Tab. \ref{tab3}). Since the stationary (within a box of size $l \ll \xi$) HDs are symmetric in $d=1+1$, MAHDs and MIHDs are identical. We recall that in \cite{HHTake2015} it was claimed that the Majumdar-Comtet (M-C) \cite{majumdar04} pdf - for (global) extremal heights of steady state $1+1$ EW interfaces with free BC - also represents the local MAHDs of $1+1$ KPZ systems. Indeed, our estimates for $S$ and $K$ are just slight smaller than the ones for the M-C distribution ($R=2.98$, $S=1.11$ and $K=1.69$), while $R$ agree within the error bar.

\begin{table}[!b]
\begin{center}
\begin{tabular}{cccccccc}
\hline\hline
dimension      &  &     $R$   &  &  $S$   &  &  $K$   & \\
\hline
$1+1$          &  &  $3.00(5)$   &  &  $1.07(2)$   &  &  $1.60(3)$  & \\
$2+1$-MAHDs    &  &  $7.3(4)$    &  &  $0.84(2)$   &  &  $1.14(5)$  & \\
$2+1$-MIHDs    &  &  $7.3(4)$    &  &  $0.93(3)$   &  &  $1.35(10)$  & \\
\hline\hline
\end{tabular}
\caption{Cumulant ratios for KPZ LEHDs.}
\label{tab3}
\end{center}
\end{table}

\begin{figure}[t]
\includegraphics[width=8.cm]{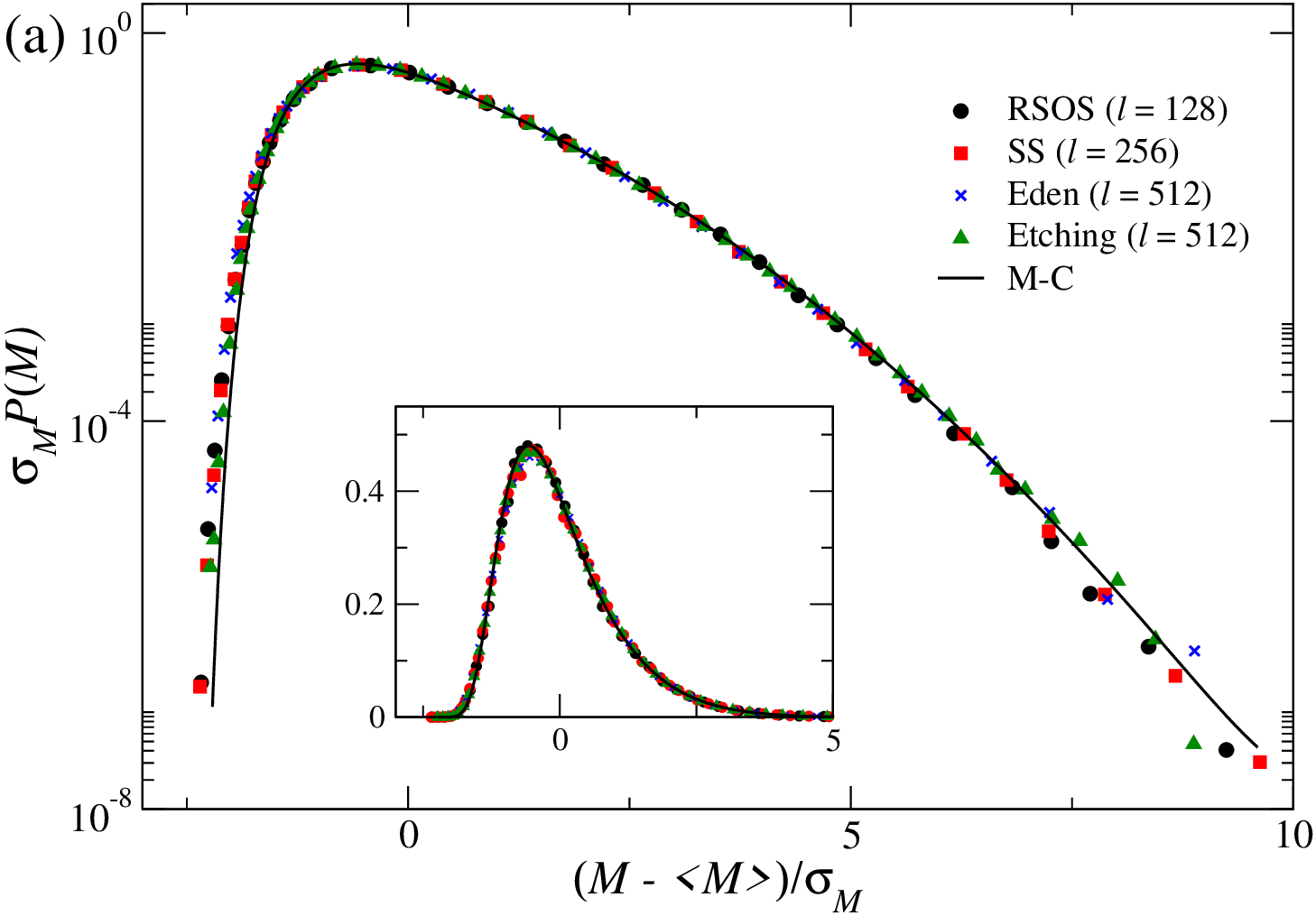}
\includegraphics[width=8.cm]{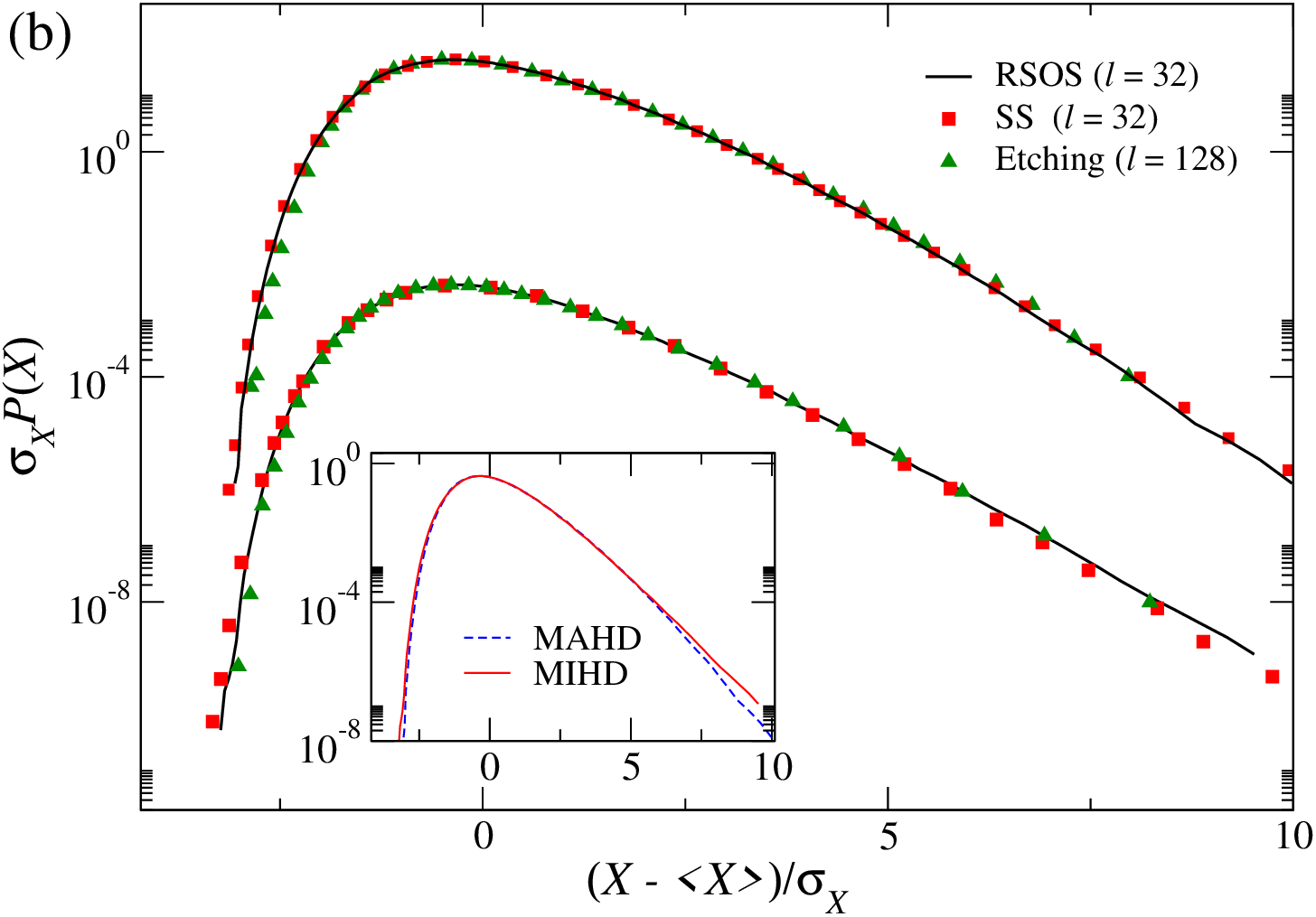}
\caption{(Color online) (a) Rescaled LEHDs for KPZ models in $d=1+1$. Inset shows the same data in linear scale. (b) Rescaled MAHDs ($X=M$, top) and MIHDs ($X=m$, bottom) for KPZ models in $d=2+1$. MAHDs (MIHDs) were shifted two decades up (down). The non-shifted MAHD and MIHD of the RSOS model are displayed in the inset. Data for fixed size (RSOS and Etching) and enlarging substrates (SS) are shown.}
\label{fig10}
\end{figure}

For $d=2+1$, the stationary KPZ HD is asymmetric and, so, different LEHDs are expected. Indeed, different values for $S$ and $K$ are found for MA- and MIHDs for RSOS/SS models (see Tab. \ref{tab3}). The MAHDs' ratios for the Etching model extrapolate to almost the same values, while the ones for MIHDs can not be extrapolated, due to a non-monotonic behavior (see Fig. \ref{fig9}d). It is worth to mention that in Ref. \cite{healy14} $S=0.884/0.877$ and $K=1.20/1.17$ were reported for MA/MIHDs for the Euler integration of KPZ equation, which are quite close our estimates in most cases.

Figures \ref{fig10}a and \ref{fig10}b show the rescaled LEHDs for $d=1+1$ and $2+1$, respectively. In the one-dimensional case, an excellent data collapse for all models and a nice agreement with the M-C pdf in the peaks and right tails are observed, but a slight difference exits in the left tail, as also found in \cite{HHTake2015}. In $d=2+1$, again, MAHDs (and MIHDs) for different models collapse quite well. As expected, from their similar cumulant ratios (Tab. \ref{tab3}), rescaled (2+1) KPZ MAHDs and MIHDs are very similar, presenting some difference only in their right tail (see inset of Fig. \ref{fig10}b).

\section{Final discussions and Conclusion}
\label{secConcl}

In summary, we have presented a detailed numerical analysis of experimentally relevant local distributions of squared roughness (SLRDs) and extremal heights (LEHDs) - calculated in the growth regime of several models in one- and two-dimensions. 

Strong finite-size effects were found in the distributions of some KPZ models, but strong evidences of their universality was obtained after appropriate extrapolations of their cumulant ratios. We claim that the procedure devised here advances over previous analysis of local distributions, since reliable estimates of cumulant ratios are obtained for relatively short times (so that $\xi \sim l$), instead of long times ($\xi \gg l$). This can be very important in the analysis of experimental interfaces, where typically $\xi$ is small and/or in universality classes with large $z$. We also emphasize that the cumulant ratio $R$, disregarded in most of previous works on local distributions, can be very useful to decide the universality class of a given system. For instance, local distributions for the MH class in $d=2+1$ have been recently studied in \cite{Iuri15}, where $R \approx 4$, $S \approx 0.88$ and $K\approx 0.85$ were found for the MAHDs ($=$ MIHDs in this class). Comparing these values with the ones in Tab. \ref{tab3}, we see that $S$ and $K$ are close to the KPZ ones, but a remarkable difference exists in $R$.

Although the underlying height fluctuations, temporal and spatial covariances, are different for flat and curved KPZ interfaces (in the growth regime) of KPZ interfaces, we find here that SRLDs and LEHDs do not present this dependence. Indeed, within a box of size $l \ll \xi$, we obtain stationary measures of $w_2$, $M$ and $m$ and, thus, our results are showing that stationary fluctuations in curved interfaces are the same as in flat ones.

Another very important finding here was the scaling of the SLRDs' cumulants at early times. We stress that this scaling advances over other methods to calculate these exponents because it is not necessary to grow the interface until the steady-state (which, generally, demands long growth/simulation times) to obtain $\alpha$ and $z$. Moreover, this (temporal) scaling seems do not suffer from crossover effects and is not affected by scaling anomalies, as does the local roughness scaling with the box size $l$. Another advantage of this method is that smooth curves of $\left\langle w_2^n \right\rangle_c \times t^{\gamma_n}$ can be obtained even for a small number of surfaces (samples), since the cumulants are averaged over several boxes at surface. Thus, we believe that this method will be very useful in experimental studies. Furthermore, it can also be important to solve theoretical/numerical issues as, for example, the KPZ exponents in higher dimensions and its related upper critical dimension.

\acknowledgments

This work is supported in part by CNPq, CAPES and FAPEMIG (Brazilian agencies).

\appendix

\section{Cumulants of the SLRD of EW class in $d=1+1$}
\label{ApEW}

Antal and R\'acz \cite{racz4} have calculated the \textit{global} roughness distribution $P(W_2,t)$ of $1+1$ EW interfaces with PBC. For flat initial conditions, they found the generating function $G_L(\zeta,t)$ of its moments as
\begin{equation}
 G_L(\zeta,t) = \mathcal{N} \prod_{m=1}^{\infty} \int d c_m d c_m^{*} \frac{\exp\left\lbrace {-2 |c_m|^2 [\zeta + 1/2 \sigma_m^2(t)] } \right\rbrace }{\sigma_m^4(t)} ,
 \label{EqApGenEW}
\end{equation}
where $\mathcal{N}$ is a normalization factor, $c_{m}$ (and $c_{-m}=c_{m}^*$) are the coefficients of the Fourier expansion of $h (x,t) - \bar{h}(t)$ and
\begin{equation}
 \sigma_m^2(t) = \frac{D (1-e^{- 2 \nu q_m^2 t})}{L \nu q_m^2},
 \label{EqApSigEW}
\end{equation}
with $q_m=2 \pi m/L$. Defining $\left\langle W_2 \right\rangle_s = D L/(12 \nu)$ (the steady state squared roughness) and $\tau \equiv 8 \pi^2 \nu t/L^2 = 4 \pi^2 [\xi(t)/L]^2$, one may write $\sigma_m^2(t) = {3 \left\langle W_2 \right\rangle_s (1-e^{- m^2 \tau})}/{(\pi m)^2}$. Calculating the Gaussian integrals, $G_L(\zeta,t)$ can be obtained as well as $\Psi_L(\zeta,t) \equiv \ln G_L(\zeta,t)$, which is the generating function of the cumulants of $P(W_2,t)$, given by
\begin{eqnarray}
 \left\langle W_2^n\right\rangle_c &=& (-1)^n \left( \frac{\partial \Psi_L}{\partial \zeta}\right) _{\zeta=0}  \\ \nonumber
  &=& \frac{(n-1)! 6^n \left\langle W_2\right\rangle_s^n}{\pi^{2 n}}  \sum_{m=1}^{\infty} \frac{(1-e^{-\tau m^2})^n}{m^{2 n}}.
\end{eqnarray}
For short times, so that $\tau \ll 1$ (and, thus, $\xi \ll L$), this sum can be approached by an integral yielding
\begin{equation}
 \left\langle W_2^n\right\rangle_c \sim \left\langle W_2\right\rangle_s^n \tau^{n-1/2} \sim (D^n/\sqrt{\nu}) L^{1-n} t^{n-1/2}.
\end{equation}

\section{SLRD of random interfaces}
\label{ApRD}

Considering a random deposition on a hypercube substrate of dimension $d_s$ and lateral size $\mathcal{L}=L a$, where $a$ is the lattice constant, $P(W_2,t)$ can be calculated by particularizing the Antal and R\'acz \cite{racz4} results, noting that

\textit{i)} $\nu=0$ in a random growth, so that the variance $\sigma_m^2(t)$ (in Eq. \ref{EqApSigEW}) have to be changed to $\sigma^2(t) = 2 D t/L^{d_s}$. Therefore, $\sigma^2$ no longer depends on the Fourier mode $m$ and, thus, all integrals in Eq. \ref{EqApGenEW} are identical; and

\textit{ii)} a cutoff have to be introduced in the product (in Eq. \ref{EqApGenEW}) of the $L^{d_s}$ modes.

After these considerations, we find the generating function
\begin{equation}
 G_L(\zeta,t)=(1+2 \zeta \sigma^2)^{-(L^{d_s}/2)}
\end{equation}
and, then, the cumulants
\begin{equation}
 \left\langle W_2^n\right\rangle_c \sim L^{d_s} \sigma^{2 n} \sim L^{(1-n)d_s} D^n t^n.
\end{equation}

Defining the adimensional variable $x=W_2/ \left\langle W_2 \right\rangle=W_2/(2 D t)$, it is easy to calculate the inverse Laplace transform
\begin{equation}
P(W_2,t) = \int_{-i \infty}^{i \infty} \frac{d \zeta}{2\pi i} G(\zeta) e^{\zeta W_2} = \frac{1}{\left\langle W_2 \right\rangle} \Phi(x,t),
\end{equation}
where
\begin{equation}
\Phi(x,t) = \frac{k^k}{(k-1)!} x^{k-1} e^{-k x},  
\end{equation}
with $k\equiv L^{d_s}/2$. Interestingly, this pdf does not depend explicitly on the time, but does on system size. Its cumulants are $\left\langle x^n \right\rangle_c = (n-1)!/k^{n-1}$. Then, when the system size diverges ($k \rightarrow\infty$), $\left\langle x^n \right\rangle_c \rightarrow 0$ for $n>1$, while $\left\langle x \right\rangle_c = 1$, i. e., the distribution tends to a delta function. By the way, notwithstanding, it approaches a log-normal distribution. In fact, performing an expansion of $\Phi$ around its mean, defining $\epsilon = x-1$ and $\sigma_x^2 \equiv \left\langle x^2 \right\rangle_c = 1/k$, we may write
\begin{equation}
\Phi(x,t) = \left( \frac{k^k e^{-k}}{(k-1)!} \right)  \exp \left[- \frac{\epsilon - \ln(1+\epsilon)}{\sigma_x^2} \right]/x,  
\end{equation}
whose term into parenthesis converges to $1/(\sqrt{2 \pi}\sigma_x)$ for large $k$, while $[\epsilon - \ln(1+\epsilon)] \approx \ln^2(x)/2 + \epsilon^3/6$ for $\epsilon \ll 1$. Then,
\begin{equation}
\Phi(x,t) \simeq  \frac{1}{\sqrt{2 \pi} \sigma_x x} {e^{- \frac{\ln^2(x)}{2 \sigma_x^2}}} \left( 1 - \frac{\epsilon^3}{6 \sigma_x^2} \right) .
\end{equation}
Defining the distance from the mean in terms of the standard deviation $\epsilon \equiv b \sigma_x$, we see that the correction $b^3/(6\sqrt{k})$ vanishes for large $k$. We remark that the cumulants of this log-normal have the form $\left\langle x \right\rangle_c = e^{\sigma_x^2/2}$, and $\left\langle x^n \right\rangle_c = \left\langle x \right\rangle_c^n (e^{\sigma_x^2} -1)^{n-1} g_n$, for $n>1$, with $g_n= const. + \mathcal{O}(e^{\sigma^2})$. Thence, if $\sigma_x \sim 1/L^{d_s}$ one finds $\left\langle x \right\rangle_c = 1$ and $\left\langle x^n \right\rangle_c \sim L^{(1-n)d_s}$.

\bibliography{KPZDist}

\end{document}